\documentclass{article}
\usepackage{amsmath,amsfonts,amssymb,amsthm,xspace,pstricks}
\newlength{\dinwidth}
\newlength{\dinmargin}
\makeatletter
\newif\if@fewtab\@fewtabtrue
\@addtoreset{equation}{section}
\def\draftdate{\number\day.\number\month.\number\year\ \ \ \hourmin }
{\count255=\time\divide\count255 by 60
\xdef\hourmin{\number\count255}
\multiply\count255 by-60\advance\count255 by\time
\xdef\hourmin{\hourmin:\ifnum\count255<10 0\fi\the\count255}}
\def\ps@draft{\let\@mkboth\@gobbletwo
    \def\@oddhead{}
    \def\@oddfoot
       {\hbox to 7 cm{$\scriptstyle\bf Draft\ version:\ \draftdate$
       \hfil}\hskip -7cm\hfil\rm\thepage \hfil}
    \def\@evenhead{}\let\@evenfoot\@oddfoot}
\makeatother
\newcommand{\Field}[1]{\ensuremath{\mathbb{#1}}\xspace}

\newcommand{\Rn}{\Field{R}}

\newcommand{\Zn}{\Field{Z}}

\newtheorem{Prp}{Proposition}
\newtheorem{Thm}{Theorem}
\newtheorem{Cor}{Corollary}
\newtheorem{Cjt}{Conjecture}
\newtheorem{Def}{Definition}

\newcommand{\lb}[1]{\label{#1}}
\newcommand{\Eq}[1]{(\ref{#1})}
\newcommand{\ct}[1]{\cite{#1}}

\renewcommand{\[}{\begin{eqnarray}}
\renewcommand{\]}{\end{eqnarray}}
\newcommand{\nn}{\nonumber}
\newcommand{\non}{\nonumber \\ }

\newcommand{\een}{\end{enumerate}}
\newcommand{\ben}{\begin{enumerate}}
\newcommand{\ga}{\alpha}

\newcommand{\gd}{\delta}
\newcommand{\gD}{\ensuremath{\Delta}\xspace}
\newcommand{\gep}{\epsilon}
\newcommand{\gf}{\varphi}
\newcommand{\gG}{\Gamma}
\newcommand{\gl}{\lambda}
\newcommand{\gL}{\ensuremath{\Lambda}\xspace}
\newcommand{\gm}{\mu}

\newcommand{\go}{\omega}

\newcommand{\gp}{\psi}

\newcommand{\gr}{\rho}

\newcommand{\gt}{\theta}
\newcommand{\gx}{\xi}

\newcommand{\cB}{\mathcal{B}}
\newcommand{\cC}{\mathcal{C}}

\newcommand{\cF}{\ensuremath{\mathcal{F}}\xspace}

\newcommand{\cH}{\mathcal{H}}
\newcommand{\cJ}{\ensuremath{\mathcal{J}}\xspace}

\newcommand{\cl}{\ensuremath{\ell}\xspace}
\newcommand{\cM}{\mathcal{M}}
\newcommand{\cN}{\mathcal{N}}
\newcommand{\cP}{\ensuremath{\mathcal{P}}\xspace}
\newcommand{\cR}{\ensuremath{\mathcal{R}}\xspace}

\newcommand{\cV}{\mathcal{V}}

\newcommand{\fg}{\ensuremath{\mathfrak{g}}\xspace}
\newcommand{\fh}{\ensuremath{\mathfrak{h}}\xspace}
\newcommand{\fn}{\mathfrak{n}}

\newcommand{\fsl}[1]{\ensuremath{\mathfrak{sl}_{#1}}\xspace}

\newcommand{\fU}{\mathfrak{U}}
\newcommand{\fw}{\mathfrak{w}}
\newcommand{\fW}{\mathfrak{W}}

\renewcommand{\vec}[1]{{\boldsymbol{#1}}}
\newcommand{\vO}{\vec{0}}
\newcommand{\1}{\vec{1}}
\newcommand{\va}{\vec{a}}

\newcommand{\vk}{\vec{k}}
\newcommand{\vl}{\vec{l}}
\newcommand{\vp}{\vec{p}}

\newcommand{\vq}{\vec{q}}

\newcommand{\vr}{\vec{r}}
\newcommand{\vrm}{{\vec{r}}_{-1}}
\newcommand{\vs}{\vec{s}}
\newcommand{\vt}{\vec{t}}

\newcommand{\vv}{\vec{v}}
\newcommand{\vx}{\vec{x}}

\newcommand{\vga}{\vec{\ga}}

\newcommand{\vgd}{\vec{\gd}}
\newcommand{\vgl}{\vec{\gl}}
\newcommand{\vgL}{\vec{\gL}}
\newcommand{\vgo}{\vec{\go}}
\newcommand{\vgr}{\vec{\gr}}

\newcommand{\vgx}{\vec{\gx}}

\DeclareMathOperator{\res}{Res}
\DeclareMathOperator{\rank}{rank}
\DeclareMathOperator{\sgn}{sgn}
\DeclareMathOperator{\End}{End}
\DeclareMathOperator{\Vir}{Vir}

\newcommand{\Span}{\mathrm{span}}
\newcommand{\mult}{\mathrm{mult}}
\newcommand{\real}{{\mathrm{re}}}
\newcommand{\imag}{{\mathrm{im}}}
\newcommand{\id}{\mathrm{id}}
\DeclareMathOperator{\ad}{ad}
\newcommand{\re}{\mathrm{e}}
\newcommand{\Res}[2][z]{\res_{#1}\left[#2\right]}
\newcommand{\frc}[2]{\tfrac{#1}{#2}}
\newcommand{\8}{\ensuremath{E_8}\xspace}
\newcommand{\9}{\ensuremath{E_9}\xspace}
\newcommand{\0}{\ensuremath{E_{10}}\xspace}
\newcommand{\2}{\tfrac12}

\newcommand{\dz}{\frac{d}{dz}}
\renewcommand{\|}{\,|\,}
\renewcommand{\.}{\cdot}
\newcommand{\X}{\!\cdot\!}

\newcommand{\ket}[1]{\ensuremath{|#1\rangle}\xspace}
\renewcommand{\:}{\boldsymbol{:}}
\newcommand{\pord}[1]{\: #1 \:}
\newcommand{\xord}[1]{{}_\times^\times #1 {}_\times^\times}

\newcommand{\cofo}[1]{\ensuremath{\langle #1 \rangle}\xspace}
\newcommand{\info}[1]{\ensuremath{( #1 )}\xspace}
\newcommand{\lico}[1]{\ensuremath{\langle #1 \rangle}\xspace}
\newcommand{\latt}[1]{\ensuremath{I\hspace{-.2em}I_{#1,1}}\xspace}
\newcommand{\fgA}{\ensuremath{\fg(A)}\xspace}
\newcommand{\fhA}{\ensuremath{\fh(A)}\xspace}
\newcommand{\fgB}{\ensuremath{\fg(B)}\xspace}

\newcommand{\fgL}{\ensuremath{\fg_\gL}\xspace}
\newcommand{\fhL}{\ensuremath{\fh_\gL}\xspace}
\newcommand{\hgA}{\ensuremath{\hat{\fg}(A)}\xspace}
\newcommand{\gDL}{\ensuremath{\gD_\gL}\xspace}
\newcommand{\fgn}[1]{\ensuremath{\fg^{[#1]}}\xspace}
\newcommand{\Lie}[1]{\ensuremath{\fg_{\latt{#1}}}\xspace}

\newcommand{\cHl}[1][\ell]{\ensuremath{{\mathcal{H}}^{[#1]}}\xspace}

\newcommand{\vkl}{\ensuremath{\vk_\cl}\xspace}
\newcommand{\AI}[2]{\ensuremath{A^{#1}_{#2}}}
\newcommand{\Ai}[1]{\ensuremath{A^i_{#1}}}
\newcommand{\Aj}[1]{\ensuremath{A^j_{#1}}}

\newcommand{\AL}[1]{\ensuremath{A^-_{#1}}}

\newcommand{\com}[1]{\bigl[ #1 \bigr]}

\begin{document}
\def\draft{\pagestyle{draft}\thispagestyle{draft}
\global\def\draftcontrol{1}}
\global\def\draftcontrol{0}
\arraycolsep3pt

\thispagestyle{empty}
\begin{flushright} hep-th/9703084
               \\  KCL-MTH-97-22
               \\  IASSNS-HEP-97/20
               \\  PSU-TH-178
               \\  AEI-029
\end{flushright}
\vspace*{2cm}
\begin{center}
 {\LARGE \sc Missing Modules, the Gnome Lie Algebra, and \0%
  }\\
 \vspace*{1cm}
 {\sl
     O.~B\"arwald\footnotemark[1]$^,$%
      \footnote[2]{Supported by \emph{Gottlieb Daimer- und Karl
                   Benz-Stiftung} under Contract No.\ 02-22/96},
     R.~W.~Gebert\footnotemark[3]$^,$%
      \footnote[4]{Supported by \emph{Deutsche Forschungsgemeinschaft}
                   under Contract No.\ DFG Ge 963/1-1},
     M.~G\"unaydin\footnotemark[5]$^,$%
      \footnote[6]{Work supported in part by \emph{NATO Collaborative
                   Research Grant} CRG.960188}$^,$%
      \footnote[7]{Work supported in part by the \emph{National
                   Science Foundation} under Grant Number PHY-9631332},
     H. Nicolai\footnotemark[8]$^,$\footnotemark[6]}\\
 \vspace*{6mm}
     \footnotemark[1]
     Department of Mathematics, King's College London\\
     Strand, London WC2R 2LS, Great Britain\\
 \vspace*{3mm}
     \footnotemark[3]
     Institute for Advanced Study, School of Natural Sciences\\
     Olden Lane, Princeton, NJ 08540, U.S.A.\\
 \vspace*{3mm}
     \footnotemark[5]
     Physics Department, Pennsylvania State University\\
     104 Davey Lab, University Park, PA 16802, U.S.A.\\
 \vspace*{3mm}
     \footnotemark[8]
     Max-Planck-Institut f\"ur Gravitationsphysik,
     Albert-Einstein-Institut \\
     Schlaatzweg 1, D-14473 Potsdam, Germany \\
 \vspace*{6mm}
\ifnum\draftcontrol=1{\Large\bf Draft version: \draftdate \\}
                     \else{11 March 1997 \\}\fi
 \vspace*{1cm}
\begin{minipage}{11cm}\footnotesize
\textbf{Abstract:}
We study the embedding of Kac--Moody algebras into Borcherds (or
generalized Kac--Moody) algebras which can be explicitly realized as
Lie algebras of physical states of some completely compactified
bosonic string. The extra ``missing states'' can be decomposed into
irreducible highest or lowest weight ``missing modules'' w.r.t.\ the
relevant Kac--Moody subalgebra; the corresponding lowest weights are
associated with imaginary simple roots whose multiplicities can be
simply understood in terms of certain polarization states of the
associated string. We analyse in detail two examples where the
momentum lattice of the string is given by the unique even unimodular
Lorentzian lattice $\latt1$ or $\latt9$, respectively. The former
leads to the Borcherds algebra $\Lie1$, which we call ``gnome Lie
algebra", with maximal Kac--Moody subalgebra $A_1$. By the use of the
denominator formula a complete set of imaginary simple roots can be
exhibited, whereas the DDF construction provides an explicit Lie
algebra basis in terms of purely longitudinal states of the
compactified string in two dimensions. The second example is the
Borcherds algebra $\Lie9$, whose maximal Kac--Moody subalgebra is the
hyperbolic algebra $E_{10}$. The imaginary simple roots at level 1,
which give rise to irreducible lowest weight modules for $E_{10}$, can
be completely characterized; furthermore, our explicit analysis of two
non-trivial level-2 root spaces leads us to conjecture that these are
in fact the only imaginary simple roots for $\Lie9$.
\end{minipage}
\end{center}
\setcounter{footnote}{0}
\newpage

\tableofcontents

\section{Introduction} \lb{sec:Int}
The main focus of this paper is the interplay between Borcherds
algebras\footnote{In the literature, these algebras are also referred
to as ``generalized Kac--Moody algebras''.} and their maximal
Kac--Moody subalgebras. The potential importance of these infinite
dimensional Lie algebras for string unification is widely recognized,
but it is far from clear at this time what their ultimate role will
be in the scheme of things (see e.g. \ct{GiPoRa94,Schwa96} for recent
overviews and motivation). In addition to their uncertain status with
regard to physical applications, these algebras are very incompletely
understood and present numerous challenges from the purely mathematical
point of view. Because recent advances in string theory have greatly
contributed to clarifying some of their mathematical intricacies
we believe that the best strategy for making progress is to exploit
string technology as far as it can take us. This is the path we will
follow in this paper.

As is well known, Kac--Moody and Borcherds algebras can both be
defined recursively in terms of a Cartan matrix $A$ (with matrix
entries $a_{ij}$) and a set of generating elements $\{e_i, f_i, h_i|
i\in I\}$ called Chevalley--Serre generators, which are subject to
certain relations involving $a_{ij}$ (see e.g.\
\ct{Kac90,MooPia95}). For Kac--Moody algebras, the matrix $A$ has to
satisfy the properties listed on page one of \ct{Kac90}; the resulting
Lie algebra is designated as $\fgA$.\footnote{We use the labeling
$i,j\in\{1,\ldots,d\}$ for $A>0$ and $i,j \in\{-1,0,1,\ldots,d-2\}$
for Lorentzian $A$ (which have Lorentzian signature), where $d=
\rank(A)$. The affine case of positive semi-definite $A$ which has a
slighly different labeling will not concern us here.} For Borcherds
algebras more general matrices $A$ are possible \ct{Borc88}; in
particular, imaginary (i.e., light-like or time-like) simple roots are
allowed, corresponding to zero or negative entries on the diagonal of
the Cartan matrix, respectively. The root system of a Kac--Moody
algebra is simple to describe, yet for any other but positive or
positive semi-definite Cartan matrices (corresponding to finite and
affine Lie algebras, resp.), the structure of the algebra itself is
exceedingly complicated and not completely known even for a single
example. By contrast, Borcherds algebras can sometimes be explicitly
realized as Lie algebras of physical states of some compactified
bosonic string. Famous examples are the fake monster Lie algebra
$\Lie{25}$ and the (true) monster Lie algebra $\fg^\natural$, arising
as the Lie algebra of transversal states of a bosonic string in 26
dimensions fully compactified on a torus or a $\Zn_2$-orbifold thereof,
respectively \ct{Borc90,Borc92}. Recently, such algebras were also
discovered in vertex operator algebras associated with the
compactified heterotic string \ct{HarMoo96}; likewise, the Borcherds
superalgebras constructed in \ct{GriNik95} may admit such explicit
realizations. However, the root systems are now much more difficult to
characterize, because one is confronted with an (generically) infinite
tower of imaginary simple roots; in fact, the full system of simple
roots is known only in some special cases.

In this paper we exploit the complementarity of these difficulties.
As shown some time ago, both Lorentzian Kac--Moody algebras and
Borcherds algebras can be conveniently and explicitly represented in
terms of a DDF construction \ct{DeDiFu72,Brow72} adapted to the root
lattice in question \ct{GebNic95a}. More precisely, any Lorentzian
algebra $\fgA$ can be embedded into a possibly larger, but in some
sense simpler Borcherds algebra of physical states $\fgL$ associated
with the root lattice $\gL$ of $\fgA$. The DDF construction then
provides a \emph{complete basis} for $\fgL$ and in particular also for
$\fgA$, although it is very difficult to determine the latter. A
distinctive feature of Lorentzian Kac--Moody algebras of
``subcritical" rank (i.e., $d<26$) is the occurrence of longitudinal
states besides the transversal ones.  This result applies in
particular to the maximally extended hyperbolic algebra \0 which can
be embedded into $\Lie9$, the Lie algebra of physical states of a
subcritical bosonic string fully compactified on the unique
10-dimensional even unimodular Lorentzian lattice \latt9. The problem
of understanding \0 can thus be reduced to the problem of
characterizing the ``missing states" (alias ``decoupled states"),
i.e.\ those physical states in $\Lie9$ not belonging to $\0$. The
problem of counting these states, in turn, is equivalent to the one of
identifying all the imaginary simple roots of $\Lie9$ with their
multiplicities.

In general terms, our proposal is therefore to study the embedding
$$
\fgA \subset \fgL,
$$
and to group the missing states
$$
\cM \equiv \fgL\ominus\fgA
$$
into an infinite direct sum of ``missing modules", that is,
irreducible highest or lowest weight representations of the subalgebra
$\fgA$. This idea of decomposing a Borcherds algebra with respect to
its maximal Kac--Moody subalgebra was already used by Kang \ct{Kang94a}
for deriving formulas for the root multiplicites of Borcherds algebras
and was treated in the axiomatic setup in great detail by Jurisich
\ct{Juri97}. We present here an alternative approach exploiting
special features of the string model. After exposing the general
structure of the embedding, we will work out two examples in great
detail. The first is $\Lie1$, the Lie algebra of physical states of a
bosonic string compactified on $\latt1$; because of its kinship with
the monster Lie algebra $\fg^\natural$ which has the same root
lattice, we will refer to it as the ``gnome Lie algebra". Its maximal
Kac--Moody subalgebra $\fgA\subset\Lie1$ is just the finite Lie
algebra $A_1\equiv \fsl2$. The other example which we will investigate
is $\Lie9$ with the maximal Kac--Moody subalgebra $\0\subset
\Lie9$. Very little is known about this hyperbolic Lie algebra, and
even less is known about its representation theory (see, however,
\ct{FeFrRi93} for some recent results on the representations of
hyperbolic Kac--Moody algebras). Our main point is that by combining
the ill-understood Lie algebra with its representations into the Lie
algebra $\Lie9$, we arrive at a structure which can be handled much
more easily.

The gnome Lie algebra has not yet appeared in the literature so far,
although it is possibly the simplest non-trivial example of a
Borcherds algebra for which not only one has a satisfactory
understanding of the imaginary simple roots, but also a completely
explicit realization of the algebra itself in terms of physical string
states. (Readers should keep in mind, that so far most investigations
of such algebras are limited to counting dimensions of root spaces and
studying the modular properties of the associated partition
functions).  It is almost ``purely Borcherds" since it has only two
real roots (and hence only one real simple root), but infinitely many
imaginary (in fact, time-like) simple roots. From the generalized
denominator formula we shall derive a generating function for their
multiplicities. Even better, the root spaces --- and not just their
dimensions --- can be analyzed in a completely explicit manner using
the DDF construction. If the fake monster Lie algebra is extremal in
the sense that it contains only transversal, but no longitudinal
states, the gnome Lie algebra $\Lie1$ is at the extreme opposite end
of the classification in that it has only longitudinal but no
transversal states. This is of course in accordance with expectations
for a $d=2$ subcritical string. Hence the gnome Lie algebra represents
the third example of a Borcherds algebra (besides the fake and the
true monster Lie algebra), for which a complete set of simple roots is
known and an explicit Lie algebra basis can be constructed.

For the Borcherds algebra \Lie9 the analysis is not so
straightforward. It has to be performed level by level where `level'
refers to the $\Zn$-grading of the Lie algebra induced by the
eigenvalue of the central element of the affine subalgebra \9 (which
makes up the level-0 piece). At level 1, we exhibit a complete set of
missing lowest weight vectors for the hyperbolic Lie algebra \0
obtainable from the tachyonic groundstate $\ket{\vr_{-1}}$ (associated
with the overextended real simple root $\vr_{-1}$) by repeated
application of the longitudinal DDF operators. To the best of our
knowledge, the corresponding \0-modules provide the first examples for
explicit realizations of unitary irreducible highest weight
representations of a hyperbolic Kac--Moody algebra. We also examine
the non-trivial root spaces associated with the two level-2 roots (or
fundamental weights w.r.t.\ the affine subalgebra) $\vgL_7$ and
$\vgL_1$, which were recently worked out explicitly in
\ct{GebNic95a,BaeGeb97} and which exemplify the rapidly increasing
complications at higher level.
An important result of this paper is the explicit
demonstration that the missing states for $\vgL_7$ and $\vgL_1$ can be
completely reproduced by commuting missing level-1 states either with
themselves or with other level-1 \0 elements. This calculation not
only furnishes a non-trivial check on our previous results, which were
obtained in a rather different manner; even more importantly, it shows
that the simple multiplicity (i.e., the multiplicity as a simple root)
of both $\vgL_7$ and $\vgL_1$ is zero. In view of this surprising
conclusion and the fact that \0 is a ``huge'' subalgebra of $\Lie9$,
we conjecture that {\em all} missing states of \0 should be obtainable
in this way.  In other words, the ``easy" imaginary simple roots of
$\Lie9$ at level-1 would in fact be the only ones. In spite of the
formidable difficulties of verifying (or falsifying) this conjecture
at arbitrary levels, we believe that its elucidation would take us a
long way towards understanding \0 and what is so special about it.

\section{The Lie Algebra of Physical States} \lb{sec:Lie}
We shall study one chiral sector of a closed bosonic string moving on
a Minkowski torus as spacetime, i.e., with all target space
coordinates compactified. Uniqueness of the quantum mechanical wave
function then forces the center of mass momenta of the string to form
a lattice \gL with Minkowskian signature. Upon ``old'' covariant
quantization this system turns out to realize a mathematical structure
called vertex algebra \cite{Borc86}. In these models the physical
string states form an infinite-dimensional Lie algebra \fgL which has
the structure of a so-called Borcherds algebra. It is possible to
identify a maximal Kac--Moody subalgebra \fgA inside \fgL which is
generically of Lorentzian indefinite type. The physical states not
belonging to \fgA are called missing states and can be grouped into
irreducible highest or lowest weight representations of \fgA. In
principle, the DDF construction allows us to identify the
corresponding vacuum states.

\subsection{The completely compactified bosonic string}
For a detailed account of this topic the reader may wish to consult
the review \cite{Gebe93}. Here, we will follow closely
\cite{GebNic95a}, omitting most of the technical details.

Let \gL be an even Lorentzian lattice of rank $d<\infty$, representing
the lattice of allowed center-of-mass momenta for the string. To each
lattice point we assign a groundstate \ket\vr which plays the role of
a highest weight vector for a $d$-fold Heisenberg algebra
$\hat{\vec{h}}$ of string oscillators $\ga_m^\mu$ ($n\in\Zn$,
$0\le\mu\le d-1$),
$$
\ga^\mu_0\ket\vr=r^\mu\ket\vr, \qquad
\ga^\mu_m\ket\vr=0\quad \forall m>0,
$$
where
$$
[\ga^\mu_m,\ga^\nu_n]=m\eta^{\mu\nu}\delta_{m+n,0}.
$$
The Fock space is obtained by collecting the irreducible
$\hat{\vec{h}}$-modules built on all possible groundstates, viz.
$$
\cF := \bigoplus_{\vr\in\gL}\cF^{(\vr)},
$$
where
$$
\cF^{(\vr)}
 :=\Span\{\ga^{\mu_1}_{-m_1}\dotsm\ga^{\mu_M}_{-m_M}\ket\vr\|
          0 \le\mu_i\le d-1,\ m_i>0\}.
$$

To each state $\gp\in\cF$, one assigns a vertex operator
$$
\cV(\gp,z)=\sum_{n\in\Zn}\gp_nz^{-n-1},
$$
which is an operator-valued ($\gp_n\in\End\cF\ \forall n$) formal
Laurent series. For notational convenience we put
$\vgx(m)\equiv\vgx\X\vga_m$ for any $\vgx\in\Rn^{d-1,1}$, and we
introduce the current
$$
\vgx(z):=\sum_{m\in\Zn}\vgx(m)z^{-m-1}.
$$
The vertex operator associated with a single oscillator is defined as
\[
\cV\bigl(\vgx(-m)\ket\vO,z\bigr)
 :=\frac{1}{(m-1)!}\left(\dz\right)^{m-1}\vgx(z),
\]
whereas for a groundstate \ket\vr one puts
\[
\cV\bigl(\ket\vr,z\bigr)
 :=\re^{\int\vr_-(z)dz}\re^{i\vr\.\vq}z^{\vr\.\vp}
   \re^{\int\vr_+(z)dz}c_{\vr},
\lb{tach-vertop} \]
with $c_{\vr}$ denoting some cocycle factor, $\vr_\pm(z):=\sum_{m>0}
\vr(\pm m) z^{\mp m-1}$, and $q^\mu$ being the position operators
conjugate to the momentum operators $p^\mu\equiv\ga^\mu_0$
($[q^\mu,p^\nu]= i\eta^{\mu\nu}$). For a general homogeneous element
$\gp=\vgx_1(-m_1)\dotsm\vgx_M(-m_M)\ket\vr$, say, the associated
vertex operator is then defined by the normal-ordered product
\[
\cV(\gp,z)
 := \;\pord{\cV\bigl(\vgx_1(-m_1)\ket\vO,z\bigr)\dotsm
         \cV\bigl(\vgx_M(-m_M)\ket\vO,z\bigr)
         \cV\bigl(\ket\vr,z\bigr)}.
\]
This definition can be extended by linearity to the whole of \cF.

The above data indeed fulfill all the requirements of a vertex algebra
\cite{Borc86,FLM88}. The two preferred elements in \cF, namely the
vacuum and the conformal vector, are given here by $\1:=\ket\vO$ and
$\vgo:=\tfrac12\vga_{-1}\X\vga_{-1}\ket\vO$, respectively. Note that
the corresponding vertex operators are respectively given by the
identity $\id_\cF$ and the stress--energy tensor
$\cV(\vgo,z)=\sum_{n\in\Zn} L_nz^{-n-2}$, where the latter provides
the generators $L_n$ of the constraint Virasoro algebra $\Vir_L$ (with
central charge $c=d$), such that the grading of \cF is obtained by the
eigenvalues of $L_0$ and the role of a translation generator is played
by $L_{-1}$ satisfying $\cV(L_{-1}\gp,z)=\dz\cV(\gp,z)$. Finally, we
mention that among the axioms of a vertex algebra there is a crucial
identity relating products and iterates of vertex operators called the
Cauchy--Jacobi identity.

We denote by $\cP^h$ the space of (conformal) highest weight vectors or
primary states of weight $h\in\Zn$, satisfying
\begin{subequations} \lb{phys-cond}
\[
L_0\gp &=& h\gp, \lb{mass-shell} \\
L_n\gp &=& 0 \quad \forall n >0.
\]
\end{subequations}
We shall call the vectors in $\cP^1$ physical states from now on. The
vertex operators associated with physical states enjoy rather simple
commutation relations with the generators of $\Vir_L$. In terms of the
mode operators we have $[L_n,\gp_m]= -m\gp_{m+n}$ for $\gp\in\cP^1$. In
particular, the zero modes $\gp_0$ of physical vertex operators
commute with the Virasoro constraints and consequently map physical
states into physical states. This observation leads to the following
definition of a bilinear product on the space of physical states
\cite{Borc86}:
\[
[\gp,\gf]:=\gp_0\gf\equiv\Res{\cV(\gp,z)\gf},
\lb{Lie-brack} \]
using an obvious formal residue notation. The Cauchy--Jacobi identity
for the vertex algebra immediately ensures that the Jacobi identity
$[\gx,[\gp,\gf]]+ [\gp,[\gf,\gx]]+ [\gf,[\gx,\gp]]= 0$ always holds
(even on \cF). But the antisymmetry property turns out to be
satisfied only modulo $L_{-1}$ terms. Hence one is led to introduce
the Lie algebra of observable physical states by
\[
\fgL := \cP^1 \big/ L_{-1}\cP^0,
\lb{Lie-phys} \]
where `observable' refers to the fact that the subspace $L_{-1}\cP^0$
consists of (unobservable) null physical states, i.e., physical states
orthogonal to all physical states including themselves (w.r.t.\ the
usual string scalar product). Indeed, for $d\neq26$, $L_{-1}\cP^0$
accounts for all null physical states.

\subsection{The DDF construction}
For a detailed analysis of \fgL one requires an explicit basis. First,
one observes that the natural \fgL-gradation by momentum already
provides a root space decomposition for \fgL, viz.
$$
\fgL=\fhL\oplus\bigoplus_{\vr\in\gD}\fgL^{(\vr)},
$$
where the root space $\fgL^{(\vr)}$ consists of all observable
physical states with momentum $\vr$:
$$
\fgL^{(\vr)}:=\{\gp\in\fgL\|p^\mu\gp=r^\mu\gp\}.
$$
The set of roots, \gD, is determined by the requirement that the
roots should represent physically allowed string momenta. Hence we
have
$$
\gD\equiv\gDL
:=\{\vr\in\gL\|\vr^2\le2,\ \vr\ne\vO\}=\gD^\real\cup\gD^\imag,
$$
where we have also split the set of roots into two subsets of real and
imaginary roots which are respectively given by
$$
\gD^\real:=\{\vr\in\gD\|\vr^2>0\}, \qquad
\gD^\imag:=\{\vr\in\gD\|\vr^2\le0\}.
$$
Zero momentum is by definition not a root but is incorporated into the
$d$-dimensional Cartan subalgebra
$$
\fhL:=\bigl\{\gx(-1)\ket\vO\|\gx\in\Rn^{d-1,1}\bigr\}.
$$
Thus the task is to find a basis for each root space. This is
achieved by the so-called DDF construction \cite{DeDiFu72,Brow72}
which we will sketch.

Given a root $\vr\in\gD$, it is always possible to find a DDF
decomposition for it,
$$
\vr=\va-n\vk \qquad\text{with } n:=1-\tfrac12\vr^2,
$$
where $\va,\vk\in\Rn^{d-1,1}$ satisfy $\va^2=2$, $\va\X\vk=1$, and
$\vk^2=0$. Having fixed $\va$ and $\vk$ we choose a set of orthonormal
polarization vectors $\vgx^i\in\Rn^{d-1,1}$ ($1\le i\le d-2$) obeying
$\vgx^i\X\va= \vgx^i\X\vk= 0$. Then the transversal and longitudinal
DDF operators are respectively defined by
\[
\Ai{m}=\Ai{m}(\va,\vk)
 &:=& \Res{\cV\left(\vgx^i(-1)\ket{m\vk},z\right)}, \\
\AL{m}=\AL{m}(\va,\vk)
 &:=& \Res{-\cV\bigl(\va(-1)\ket{m\vk},z\bigr)
           +\frac{m}2\dz \log\bigl(\vk(z)\bigr)
            \cV\bigl(\ket{m\vk},z \bigr)}  \non
   && -\frac12\sum_{n\in\Zn}\xord{\Ai{n}\Ai{m-n}}
      +2\gd_{m0}\vk\X\vp. \]
We shall need to make use of the following important facts about
the DDF operators (see e.g.\ \ct{GebNic95a}).

\begin{Thm} \lb{thm-DDF}
Let $\vr\in\gD$. The DDF operators associated with the DDF
decomposition $\vr=\va-n\vk$ enjoy the following properties on the
space of physical string states with momentum $\vr$, $\cP^{1,(\vr)}$:
\ben
\item(Physicality)\quad
$[L_m,\Ai{n}]=[L_m,\AL{n}]=0$;
\item(Transversal Heisenberg algebra)\quad
$[\Ai{m},\Aj{n}]=m\gd^{ij}\gd_{m+n,0}$;
\item(Longitudinal Virasoro algebra)\quad
$[\AL{m},\AL{n}]=(m-n)\AL{m+n}+\frac{26-d}{12}m(m^2-1)\gd_{m+n,0}$;
\item(Null states)\quad
$\AL{-1}\ket\va\propto L_{-1}\ket{\va-\vk}$;
\item(Orthogonality)\quad
$[\AL{m},\Ai{n}]=0$;
\item(Highest weight property)\quad
$\Ai{k}\ket{\va}=\AL{k}\ket{\va}=0$ for all $k\ge0$;
\item(Spectrum generating)\quad
$\cP^{1,(\vr)}=\Span\bigl\{\AI{i_1}{-m_1}\dotsm\AI{i_M}{-m_M}
                    \AI{-}{-n_1}\dotsm\AI{-}{-n_N}\ket{\va}\|
                    m_1+\dots+n_N=1-\tfrac12\vr^2\bigr\}$;
\een
for all $m,n\in\Zn$ and $1\le i\le d-2$.
\end{Thm}

As a simple consequence, we have the following explicit formula for
the multiplicity of a root $\vr$ in \fgL:
\[
\mult_{\fgL}(\vr)\equiv\dim\fgL^{(\vr)}=
\pi_{d-1}(n):=p_{d-1}(n)-p_{d-1}(n-1),
\lb{genmult-1} \]
where $n=1-\tfrac12\vr^2$ and $\sum_{n\ge0}p_d(n)q^n=
[\phi(q)]^{-d}\equiv \prod_{n\ge1}(1-q^n)^{-d}$, so that
\[
\sum_{n=0}^\infty\pi_{d-1}(n)q^n=
\frac{1-q}{[\phi(q)]^{d-1}}=
1+(d-2)q+\tfrac12(d-1)dq^2+\dotsb.
\lb{genmult-2} \]

The above theorem is also useful for constructing a positive definite
symmetric bilinear form on \fgL as follows:
$$
\cofo{\vr|\vs}:=\gd_{\vr,\vs}\quad\text{for } \vr,\vs\in\gL,\qquad
(\ga_m^\mu)^\dagger:=\ga_{-m}^\mu.
$$
For the DDF operators this yields
$$
(\Ai{m})^\dagger=\Ai{-m},\qquad
(\AL{m})^\dagger=\AL{-m}.
$$
In view of the above commutation relations it is then clear that
$\cofo{\ |\ }$ is positive definite on any root space $\fgL^{(\vr)}$
if $d<26$. For the critical dimension, $d=26$, we redefine \fgL by
dividing out the additional null states which correspond to the
remaining longitudinal DDF states. Thus we have to replace $\pi_{d-1}$
by $p_{24}$ in the multiplicity formula. Note that the scalar product
has Minkowskian signature on the Cartan subalgebra.

For our purposes we shall also need an invariant symmetric bilinear
form on \fgL which is defined as
$$
\info{\gp|\gf}:=-\cofo{\gt(\gp)|\gf}
$$
for $\gp,\gf\in\fgL$, where the Chevalley involution is given by
$$
\gt(\ket\vr):=\ket{-\vr},\qquad
\gt\circ\ga^\mu_m\circ\gt:=-\ga^\mu_m.
$$
Clearly, both bilinear forms are preserved by this involution and they
enjoy the invariance and contravariance properties, respectively, viz.
\[
\info{[\gp,\chi]|\gf}=\info{\gp|[\chi,\gf]}, \quad
\cofo{[\gp,\chi]|\gf}=\cofo{\gp|[\gt(\chi),\gf]}
\quad\forall\gp,\chi,\gf\in\fgL.
\lb{incovariance} \]

\subsection{Borcherds algebras and Kac--Moody algebras} \lb{subsec:Borch}
We now have all ingredients at hand to show that \fgL for any $d>0$
belongs to a certain class of infinite-dimensional Lie algebras.

\begin{Def} \lb{def-B}
Let $J$ be a countable index set (identified with some subset of
$\Zn$). Let $B= (b_{ij})_{i,j\in J}$ be a real matrix, satisfying the
following conditions:
\ben
\item[(C1)]
  $B$ is symmetric;
\item[(C2)]
  If $i\ne j$ then $b_{ij}\le0$;
\item[(C3)]
  If $b_{ii}>0$ then $\frac{2b_{ij}}{b_{ii}}\in\Zn$ for all $j\in J$.
\een
Then the universal Borcherds algebra $\fgB$ associated with $B$ is
defined as the Lie algebra generated by elements $e_i$, $f_i$ and
$h_{ij}$ for $i,j\in J$, with the following relations:
\ben
\item[(R1)]
  $[h_{ij},e_k]=\gd_{ij}b_{ik}e_k$, \quad
  $[h_{ij},f_k]=-\gd_{ij}b_{ik}f_k$;
\item[(R2)]
  $[e_i,f_j]=h_{ij}$;
\item[(R3)]
  If $b_{ii}>0$ then
  $(\ad e_i)^{1-2b_{ij}/b_{ii}}e_j=
   (\ad f_i)^{1-2b_{ij}/b_{ii}}f_j=0$;
\item[(R4)]
  If $b_{ij}=0$ then $[e_i,e_j]=[f_i,f_j]=0$.
\een
\end{Def}

The elements $h_{ij}$ span an abelian subalgebra of $\fgB$ called the
Cartan subalgebra. In fact, the elements $h_{ij}$ with $i\ne j$ lie in
the center of $\fgB$. It is easy to see that $h_{ij}$ is zero unless the
$i$th and $j$th columns of the matrix $B$ are equal.\footnote{Actually,
the elements $h_{ij}$ for $i\neq j$ do not play any role and in fact
cannot appear in the present context, where $\fgB = \fgL$ is based on a
non-degenerate Lorenzian lattice $\gL$. Namely, for the
$i$th and $j$th columns of $B$ to be equal the corresponding roots
must be equal, and therefore such $h_{ij}$ are always of the form
${\vv}_{ij}(-1)\ket\vO$ with $\vv_{ij}\in\gL$. Since furthermore
the $h_{ij}$ with $i\ne j$ are central elements, the lattice vectors
$\vv_{ij}$ must be orthogonal to all (real and imaginary) roots.
Because $\gL$ is non-degenerate, we conclude that $\vv_{ij} = 0$, and
hence $h_{ij}= 0$ for $i\neq j$.} A Lie algebra
is called Borcherds algebra, if it can be obtained from a universal
Borcherds algebra by dividing out a subspace of its center and adding
an abelian algebra of outer derivations. An important property of
(universal) Borcherds algebras is the existence of a triangular
decomposition
\[
\fg=\fn_-\oplus\fh\oplus\fn_+,
\lb{trian-dec}
\]
where $\fn_+$ and $\fn_-$ denote the subalgebras generated by the
$e_i$'s and the $f_i$'s, respectively. This can be established by the
usual methods for Kac--Moody algebras (see \ct{Juri97} for a careful
proof).

Given the Lie algebra of physical string states, \fgL, it is extremely
difficult to decide whether it is a Borcherds algebra in the sense of
the above definition. Luckily, however, there are alternative
characterizations of Borcherds algebras which can be readily applied
to the case of \fgL. We start with the following one \ct{Borc88}.

\begin{Thm} \lb{thm-B1}
A Lie algebra \fg is a Borcherds algebra if it has an almost positive
definite contravariant form $\cofo{\ |\ }$, which means that \fg has
the following properties:
\ben
\item(Grading)\quad
  $\fg=\bigoplus_{n\in\Zn}\fg_n$ with $\dim\fg_n<\infty$ for $n\ne0$;
\item(Involution)\quad
  $\fg$ has an involution $\gt$ which acts as $-1$ on $\fg_0$ and maps
  $\fg_n$ to $\fg_{-n}$;
\item(Invariance)\quad
  $\fg$ carries a symmetric invariant bilinear form $\info{\ |\ }$
  preserved by $\gt$ and such that $\info{\fg_m\|\fg_n}=0$ unless
  $m+n=0$;
\item(Positivity)\quad
  The contravariant form $\cofo{x|y}:=-\info{\gt(x)|y}$ is positive
  definite on $\fg_n$ if $n\neq0$.
\een
\end{Thm}

The converse is almost true, which means that, apart from some
pathological cases, a Borcherds algebra always satisfies the
conditions in the above theorem (cf.\ \ct{Juri97}).

Hence \fgL for $d\le26$ is a Borcherds algebra if we can equip it with
an appropriate \Zn-grading. Note that the grading given by assigning
degree $1-\tfrac12\vr^2$ to a root space $\fgL^{(\vr)}$ will not work
since there are infinitely many lattice points lying on the
hyperboloid $\vx^2=\text{const}\in2\Zn$. The solution is to slice the
forward (resp.\ backward) light cone by a family of
$(d-1)$-dimensional parallel hyperplanes whose common normal vector is
timelike and has integer scalar product with all the roots of \fgL(
i.e., it is an element of the weight lattice $\gL^*$). There is one
subtlety here, however. It might well happen that for a certain choice
of the timelike normal vector $\vt\in\gL^*$ there are some real roots
$\vr\in\gL$ which are orthogonal to $\vt$ so that the associated root
spaces would have degree zero.\footnote{By choosing $\vt$ to be
timelike it is also assured that it has nonzero scalar product with
all imaginary roots.} But then we would run into trouble
since the Chevalley involution does not act as $-1$ on a root space
$\fgL^{(\vr)}$ but rather maps it into $\fgL^{(-\vr)}$. We call a
timelike vector $\vt\in\gL^*$ grading vector if it is ``in general
position,'' which means that it has nonzero scalar product with all
roots. So let us fix some grading vector\footnote{Grading vectors
always exist since the hyperplanes orthogonal to the real roots cannot
exhaust all the points of $\gL^*$ inside the lightcone.} and define
$$
\fg_n:=\bigoplus_{\substack{\vr\in\gD\\ \vr\.\vt=n}}\fgL^{(\vr)},\qquad
\fg_0:=\fhL
$$
(The associated degree operator is just $\vt\X\vp$.) Then this yields
the grading necessary for \fgL to be a Borcherds algebra. Note that
the pairing property $\info{\fg_m\|\fg_n}\propto \gd_{m+n,0}$ is
fulfilled since $\gt$ is induced from the reflection symmetry of the lattice.
Observe also that if the lattice admits a (time-like) Weyl vector $\vgr$
we can set $\vt = \vgr$ since this vector has all the requisite properties.
We conclude: if the lattice $\gL$ has a grading vector and
$d\le26$, then the Lie algebra of physical states, \fgL, is a
Borcherds algebra. This result suggests that above the critical
dimension the Lie algebra of physical string states somehow changes in
type, as one would also naively expect from a string theoretical point
of view. But this impression is wrong. It is an artefact caused by the
special choice of the string scalar product. To see this, we recall
another characterization of Borcherds algebras \ct{Borc95b}.

\begin{Thm} \lb{thm-B2}
A Lie algebra \fg satisfying the following conditions is a
Borcherds algebra:
\ben
\item[(B1)]
 \fg has a nonsingular invariant symmetric bilinear form $\info{\ |\ }$;
\item[(B2)]
 \fg has a self-centralizing subalgebra \fh such that \fg is
 diagonalizable with respect to \fh and all the eigenspaces are
 finite-dimensional;
\item[(B3)]
 \fh has a regular element $h^\times$, i.e., the centralizer of
 $h^\times$ is \fh and there are only a finite number of roots
 $\vr\in\fh^*$ such that $|\vr(h^\times)|<R$ for any $R\in\Rn$;
\item[(B4)]
 The norms of roots of \fg are bounded above;
\item[(B5)]
 Any two imaginary roots which are both positive or negative have
 inner product at most 0, and if they are orthogonal their root spaces
 commute.
\een
\end{Thm}

Here, as usual, the nonzero eigenvalues of \fh acting on \fg are
elements of the dual $\fh^*$ and are called roots of \fg. A root is
called positive or negative depending on whether its value on the
regular element is positive or negative, respectively; and a root is
called real if its norm (naturally induced from $\info{\ |\ }$ on \fg)
is positive, and imaginary otherwise. Note that the regular element
provides a triangular decomposition \Eq{trian-dec} by gathering all
root spaces associated with positive (resp.\ negative) roots into the
subalgebra $\fn_+$ (resp.\ $\fn_-$).

For our purposes we shall need a special case of this theorem. Suppose
that the bilinear form has Lorentzian signature on \fh (and
consequently also on $\fh^*$). For the regular element $h^\times$ we
can take any $\vt(-1)\ket\vO$ associated with a timelike vector $\vt$
in general position (cf.\ the above remark about grading vectors!). But
the Lorentzian geometry implies more; namely, that two vectors inside
or on the forward (or backward) lightcone have to have nonpositive
inner product with each other, and they can be orthogonal only if they
are multiples of the same lightlike vector. Therefore we have
\ct{Borc95b}

\begin{Cor} \lb{cor-B1}
A Lie algebra \fg satisfying the following properties conditions is a
Borcherds algebra:
\ben
\item[(B1')]
 \fg has a nonsingular invariant symmetric bilinear form $\info{\ |\ }$;
\item[(B2')]
 \fg has a self-centralizing subalgebra \fh such that \fg is
 diagonalizable with respect to \fh and all the eigenspaces are
 finite-dimensional;
\item[(B3')]
 The bilinear form restricted to \fh is Lorentzian;
\item[(B4')]
 The norms of roots of \fg are bounded above;
\item[(B5')]
 If two roots are positive multiples of the same norm 0 vector then
 their root spaces commute.
\een
\end{Cor}

Apparently, \fgL for any $d$ fulfills the conditions
\textit{(B1')--(B4')}. A straightforward exercise in oscillator
algebra also verfies \textit{(B5')} (see formula (3.1) in
\ct{GebNic95a}). We conclude that \fgL is indeed always a Borcherds
algebra. Although we do not know the Cartan matrix $B$ associated to
\fgL (and so the set of simple roots) we can determine the maximal
Kac--Moody subalgebra of \fgL given by the
submatrix $A$ obtained from $B$ by deleting all rows and columns $j\in
J$ such that $b_{jj}\le0$.

A special role is played by the lattice vectors of length 2 which are
called the real roots of the lattice and which give rise to tachyonic
physical string states. Lightlike or timelike roots are referred to as
imaginary roots. We associate with every real root $\vr\in\gL$ a
reflection by $\fw_\vr(\vx) := \vx - (\vx\X\vr) \vr$ for
$\vx\in\Rn^{d-1,1}$. The reflecting hyperplanes then divide the vector
space $\Rn^{d-1,1}$ into regions called Weyl chambers. The reflections
in the real roots of \gL generate a group called the Weyl group $\fW$
of $\gL$, which acts simply transitively on the Weyl chambers. Fixing
one chamber to be the fundamental Weyl chamber $\cC$ once and for all,
we call the real roots perpendicular to the faces of $\cC$ and with
inner product at most 0 with elements of $\cC$, the simple roots. We
denote such a set of real simple roots by $\Pi^\real= \Pi^\real(\cC)=
\{\vr_i|i\in I\}$ for a countable index set $I$.\footnote{$I$ may be
identified with a subset of $J$. Note, however, that apart from some
special examples, the matrix $B$ for \fgL as Borcherds algebra is not
known.} Note that a priori there is no relation between the rank $d$
of the lattice and the number of simple roots, $|I|$.\footnote{The
extremal case occurs for the Lattice \latt{25} where $d=26$ but
$|I|=\infty$ \ct{Conw83}. We should mention here that in order to get
the set of imaginary roots ``well-behaved,'' one assumes that the
semidirect product of the Weyl group with the group of graph
automorphisms associated with the Coxeter--Dynkin diagram of
$\Pi^\real$ has finite index in the automorphism group of the lattice
$\gL$ (see e.g.\ \cite{Niku95}).}

The main new feature of Borcherds algebras in comparison with ordinary
Kac--Moody algebras is the appearance of \emph{imaginary simple
roots}. An important property of Borcherds algebras is the existence
of a character formula which generalizes the Weyl--Kac character
formula for ordinary Kac--Moody algebras and which leads as a special
case to the following Weyl--Kac--Borcherds denominator formula.

\begin{Thm} \lb{thm-WKB}
Let \fg be a Borcherds algebra with Weyl vector $\vgr$ (i.e.,
$\vgr\X\vr= -\tfrac12\vr^2$ for all simple roots) and Weyl group $\fW$
(generated by the reflections in the real simple roots). Then
\[
\prod_{\vr\in\gD_+}(1-\re^\vr)^{\mult(\vr)}
=\sum_{\fw\in\fW}(-1)^\fw\re^{\fw(\vgr)-\vgr}
 \sum_{\vs}\gep(\vs)\re^{\fw(\vs)},
\lb{WKB} \]
where $\gep(\vs)$ is $(-1)^n$ if $\vs$ is a sum of $n$ distinct
pairwise orthogonal imaginary simple roots and zero otherwise.
\end{Thm}

Note that the Weyl vector may be replaced by any other vector having
inner product $-\tfrac12\vr^2$ with all \emph{real} simple roots since
$\re^{\fw(\vgr)-\vgr}$ involves only inner products of $\vgr$ with
real simple roots. This will be important for the gnome Lie algebra
below where there is no true Weyl vector but the denominator formula
nevertheless can be used to determine the multiplicities of the
imaginary simple roots.

The physical states
\[
e_i:=\ket{\vr_i},\qquad
f_i:=-\ket{-\vr_i},\qquad
h_i:=\vr_i(-1)\ket\vO,
\lb{gen-def} \]
for $i\in I$, obey the following commutation relations (see
\cite{Borc86}):
\begin{subequations} \lb{gen-com}
\begin{alignat}{2}
[h_i,h_j] &= 0, & & \lb{gen-com1} \\ {}
[e_i,f_j] &= \gd_{ij}h_i, & & \lb{gen-com2} \\ {}
[h_i,e_j] &= a_{ij}e_j, & [h_i,f_j] &= -a_{ij}f_j, \lb{gen-com3} \\ {}
(\ad e_i)^{1-a_{ij}}e_j &= 0, &
(\ad f_i)^{1-a_{ij}}f_j &= 0\quad\forall i\neq j, \lb{gen-com4}
\end{alignat}
\end{subequations}
which means that they generate via multiple commutators the Kac--Moody
algebra $\fg(A)$ associated with the Cartan matrix $A=
(a_{ij})_{i,j\in I}$, $a_{ij}:= \vr_i \X \vr_j$. As usual, we
have the triangular decomposition
\[
\fgA=\fn_-(A)\oplus\fhA\oplus\fn_+(A),
\lb{trian-A} \]
where $\fn_+(A)$ (resp.\ $\fn_-(A)$) denotes the subalgebra generated
by the $e_i$'s (resp.\ $f_i$'s) for $i\in I$. This corresponds to a
choice of the grading vector $\vt$ (and the regular element $h^\times
:= \vt(-1)\ket{\vO}$) satisfying $\vt\X\vr_i>0$ $\forall i\in I$.
The Lie algebra $\fg(A)$ is a proper subalgebra of the Lie algebra of
physical states $\fgL$,
$$
\fgA \subset \fgL.
$$
If we finally introduce the Kac--Moody root lattice
$$
Q(A):= \sum_{i\in I}\Zn\vr_i,
$$
then obviously $Q(A)\subseteq\gL$ and in particular $\rank
Q(A)\le d$, even though $|I|$ might be larger than $d$.

\subsection{Missing modules}
Having found the Kac--Moody algebra \fgA, the idea is now to analyze
the ``rest'' of \fgL from the point of view of \fgA. It is clear that,
via the adjoint action, \fgL is a representation of \fgA. Since the
contravariant bilinear form is positive definite on the root spaces
$\fgL^{(\vr)}$, $\vr\in\gD$, it is sensible to consider the direct sum
of orthogonal complements of $\fgA\cap\fgL^{(\vr)}$ in $\fgL^{(\vr)}$
with respect to $\cofo{\ |\ }$ and explore its properties under the
action of \fgA. We shall see that the resulting space of so-called
missing states is a completely reducible \fgA-module, decomposable
into irreducible highest or lowest weight representations. The issue
of zero momentum, however, requires some care. If $Q(A)\neq\gL$, then
there must be a set of $d-\rank Q(A)$ linearly independent imaginary
simple roots, $\{\vr_j|j\in H\subset J\setminus I\}$, linearly
independent of the set of real simple roots, such that $\fhL=
\fhA\oplus\fh'$ with $\fh':= \Span\{h_j|j\in H\}$. The latter subspace
of the Cartan subalgebra is in general not a \fgA-module but rather an
abelian algebra of outer derivations for \fgA in view of the
commutation relations \textit{(R1)}. This observation suggests to
consider an extension of \fgA by these derivations. There is also
another argument that this is a natural thing to do. Namely, extending
\fhA to \fhL ensures that any root $\vr$ is a nonzero weight for the
extended Lie algebra, while this is not guaranteed for \fgA because
there might exist roots in $\gD$ orthogonal to all real simple
roots. This procedure is in spirit the same for the general theory of
affine Lie algebras where one extends the algebra by adjoining outer
derivations to the Cartan subalgebra such that the standard invariant
form becomes nondegenerate.

\begin{Def}
The Lie algebra $\hgA := \fgA+\fhL = \fn_-(A)\oplus\fhL\oplus\fn_+(A)$
is called extended Kac--Moody algebra associated with \gL. The
orthogonal complement of \hgA in \fgL with respect to the
contravariant bilinear form $\cofo{\ |\ }$ is called the space of
missing (or decoupled) states, $\cM$.
\end{Def}

It is clear that \hgA has the same root system and root space
decomposition as \fgA. Note that $\cM$ is the same as the orthogonal
complement of \hgA in \fgL with respect to the invariant form $\info{\
|\ }$. Obviously, $\cM$ has zero intersection with the Cartan
subalgebra \fhL and with all the tachyonic root spaces $\fgL^{(\vr)}$,
$\vr\in\gD^\real=\gD^\real(A)$. Hence we can write
\[
\cM = \cM_-\oplus\cM_+, \qquad
\cM_\pm:=\bigoplus_{\vr\in\gD^{\imag}_\pm}\cM^{(\vr)},
\]
where $\gD^{\imag}_\pm$ denotes the set of imaginary roots inside the
forward or the backward lightcone, respectively,\footnote{This means
that we choose the grading vector to lie inside the backward
lightcone} and $\cM^{(\vr)}$ is given as the orthogonal complement of
the root space for \fgA in \fgL, viz.
\[
\fgL^{(\vr)}=\fgA^{(\vr)}\oplus\cM^{(\vr)}\qquad
\forall\vr\in\gD^{\imag}.
\]
Note that it might (and in some examples does) happen that
$\fgA^{(\vr)}$ is empty for some $\vr\in\gD^{\imag}$, namely when
$\vr$ is not a root for \fgA.

Generically, \fgA is a (infinite-dimensional) Lorentzian Kac--Moody
algebra about which not much is known. On the other hand we are in the
lucky situation of having a root space decomposition with known
multiplicities for \fgL. So the main problem in this string
realization of \fgA is to understand the space of missing states. The
starting point for the analysis presented below is the following
theorem \ct{Juri97}.

\begin{Thm} \lb{thm-miss}
\ben
\item
$\cM$ is completely reducible under the adjoint action of \fgA. It
decomposes into an orthogonal (w.r.t.\ \cofo{\ |\ }) direct sum of
irreducible lowest or highest weight modules for \fgA:
\[
\cM_\pm=\bigoplus_{\vr\in\cB}m_{\vr}L(\mp\vr),
\]
where $\cB\subset\gL\cap(-\cC)$ denotes some appropriate set of
dominant integral weights for \fhA, $L(\vr)$ (resp.\ $L(-\vr)$)
denotes an irreducible highest (resp.\ lowest) weight module for \fgA
with highest weight $\vr$ (resp.\ lowest weight $-\vr$), which occurs
with multiplicity $m_{\vr}$ ($=m_{-\vr}$) inside $\cM_-$ (resp.\
$\cM_+$).
\item
Let $\cH_\pm\subset\cM_\pm$ denote the space of missing lowest and
highest weight vectors, respectively. Equipped with the bracket in
\fgL, $\cH_+$ and $\cH_-$ are (isomorphic) Lie algebras. If there are
no pairwise orthogonal imaginary simple roots in \fgL, then they are
free Lie algebras.
\een
\end{Thm}

\begin{proof}
Let $x\in\hgA$, $m\in\cM$. Then we can write $[x,m]= ax'+ bm'$
for some $x'\in\hgA$, $m'\in\cM$, $a,b\in\Rn$. It follows that
$a\info{y|x'}= \info{y|[x,m]}= \info{[y,x]|m}= 0$ for all $y\in\hgA$
using invariance. Since the radical of the invariant form has been
divided out we conclude that $a=0$. Thus $[\hgA,\cM]\subseteq\cM$ and
the homomorphism property of $\gr: \fgA \to \End\cM$,
$\gr(x)m:=[x,m]$, follows from the Jacobi identity in \fgL. But $\cM_\pm$
are already \hgA-modules by themselves. To see this, we exploit the
\Zn-grading of \fgL induced by the grading vector $\vt$. An element of
\fgL with momentum $\vr$ is said to have height $\vr\X\vt$. Then
$\cM_+$ and $\cM_-$ consist of elements of positive and negative
height, respectively. Going from positive to negative weight with the
action of \hgA requires missing states of height zero, which cannot
exist since $\fhL\subset\hgA$.

By applying the Chevalley involution $\gt$, it is sufficient to
consider $\cM_-$. Let $\cN\subset\cM_-$ be a \hgA-submodule. Then
$$
\cN = \bigoplus_{\vr\in\gD^{\imag}_-}\cN^{(\vr)},\qquad
\cN^{(\vr)} := \cM_-^{(\vr)}\cap\cN.
$$
Since $\dim\cM_-^{(\vr)}\le \dim{\fgL}_-^{(\vr)}< \infty$ and $\cofo{\
|\ }$ is positive definite on $\cM_-^{(\vr)}$ for all $\vr\in\gD$, it
follows that we have the decomposition
$$
\cM_-^{(\vr)}=\cN^{(\vr)}\oplus\cN^{(\vr)\perp}\qquad
\forall \vr\in\gD^{\imag}_-.
$$
If we define
$$
\cN^\perp := \bigoplus_{\vr\in\gD^{\imag}_-}\cN^{(\vr)\perp},
$$
then
$$
\cM_-=\cN\oplus\cN^\perp,
$$
and
$$
\cofo{\cN|x(m)}=\cofo{\gt(x)(\cN)|m}=0
$$
for all $x\in\hgA$, $m\in\cN^\perp$, since $\cN$ is a submodule by
assumption. Hence $\cN^\perp$ is also a \hgA-submodule and $\cM_-$ is
indeed completely reducible.

Finally, it is easy to see that each
irreducible \hgA-submodule $\cN\subset\cM_-$ is of highest-weight
type. Indeed, $\cN$ inherits the grading of $\cM_-$ by height which is
bounded from above by zero, whereas the Chevalley generators $e_i$
($i\in I$) associated with real simple roots increase the height when
applied to elements of $\cN$.

Now we want to show that each irreducible \hgA-module
$\cN\subset\cM_-$ is also irreducible under the action of \fgA. We
shall use an argument similar to the proof of Prop.\ 11.8 in
\ct{Kac90}. Recall that we have the decomposition $\fhL= \fhA\oplus
\fh'$, where $\fh'$ is spanned by suitable elements $h_i=
\vr_i(-1)\ket\vO$ ($i\in H$) associated with imaginary simple roots
$\vr_i$. Obviously, any imaginary simple root $\vr_i$ satisfies
$\vr_i\X\vr\ge0$ for all $\vr\in\gD^{\imag}_-$ and $\vr_i\X\vr_j\le0$
for all $\vr_j\in\Pi^{\real}$. Let us introduce a restricted grading
vector by $\vt':= \sum_{i\in H}\vr_i$. We shall call the inner product
of $\vt'$ with any root $\vr$ the restricted height of that root. The
subspaces of $\cN$ of constant restricted height are then given by
$$
\cN_h:=\bigoplus_{\substack{\vr\in\gD^{\imag}_- \\ \vt'\.\vr=h}}
       \cN^{(\vr)}.
$$
Since $\vt'\X\vr\ge0$ for all $\vr\in\gD^{\imag}_-$, there exists some
minimal $h_{\text{min}}$ such that $\cN_{h_{\text{min}}}\ne0$ and
$\cN_h=0$ for $h<h_{\text{min}}$. We have a decomposition of \fgA
w.r.t.\ to the restricted height as well, viz.
$$
\fgA=\fg_-\oplus\fg_0\oplus\fg_+,\qquad
\fg_\pm:=\bigoplus_{h\gtrless0}\fgA_h.
$$
Note that this triangular decomposition is different from the previous
one encountered in \Eq{trian-A}. In general, they are related by
$\fn_\pm(A)\subseteq \fg_\pm\oplus\fg_0$ and $\fhA\subseteq \fg_0$.
Now, apparently each $\cN_h$ is a $\fg_0$ module. In particular,
$\cN_{h_{\text{min}}}$ must be irreducible, since any
$\fg_0$-invariant proper subspace would generate a proper \hgA
submodule of $\cN$ contradicting its irreducibility. By the same
argument, $\{v\in\cN_h| \fg_-(v)= 0\}= 0$ for $h> h_{\text{min}}$.
Hence
$$
\cN= \fU(\fg_+)\cN_{\text{vac}},
$$
where
$$
\cN_{\text{vac}}:= \{v\in\cN| \fg_-(v)= 0\}= \cN_{h_{\text{min}}}
$$
is an irreducible $\fg_0$ module. From this we conclude that $\cN$ is
indeed an irreducible $\fgA$ module.

So $\cM_-$ decomposes into an othogonal direct sum
$$
\cM_-=\bigoplus_{\ga\in\cB}m_{\ga}L_{\ga},
$$
where $\cB$ denotes some appropriate index set and each $L_{\ga}$ is
an irreducible \fgA-module occurring with multiplicity $m_{\ga}>0$.
Finally, it is easy to see that each irreducible \fgA-submodule
$L_{\ga}\subset\cM_-$ is of highest-weight type. Indeed, $L_{\ga}$
inherits the grading of $\cM_-$ by height which is bounded from above
by zero, whereas the Chevalley generators $e_i$ ($i\in I$) associated
with real simple roots increase the height when applied to vectors in
$L_{\ga}$. So there exists an element $v_{\vr}\in L_{\ga}$ associated
with a dominant integral weight $\vr\in \gL\cap(-\cC)$ such that
$e_i(v_{\vr})= 0$ for all $i\in I$ and $L_{\ga}\equiv L(\vr)=
\fU\bigl(\fn_-(A)\bigr)v_{\vr}$.

To prove the second part of the theorem, let $v_1,v_2\in\cH_-$. It
follows that $x\bigl([v_1,v_2]\bigr):= [x,[v_1,v_2]]= [x(v_1),v_2]+
[v_1,x(v_2)]$. If we choose $x=e_i$ or $x=h_i$, respectively, it is
clear that $[v_1,v_2]$ is again a highest weight vector. To see that
it is missing we note that $\cofo{x|[v_1,v_2]}=
\cofo{x\bigl(\gt(v_1)\bigr)|v_2}$ for all $x\in\fgA$ by
contravariance. But since $x\bigl(\gt(v_1)\bigr)\in\cM_+$ and $v_2\in
\cM_-\perp \cM_+$ we see that indeed $[v_1,v_2]\in\cH_-$. Finally,
since \fgL is a Borcherds algebra we know that extra Lie algebra
relations (in addition to those for \fgA) can occur only if there are
pairwise orthogonal imaginary simple roots in \fgL. If this is not the
case $\cH_\pm$ must be free.
\end{proof}

So the space of missing states decomposes into an orthogonal direct
sum of irreducible \fgA-multiplets each of which is obtained by
repeated application of the raising operators $e_i$ (resp.\ $f_i$) to
some lowest (resp.\ highest) weight vector. This beautiful structure,
however, looks rather messy from the point of view of a single missing
root space, $\cM^{(\vr)}$, say. Generically, it decomposes into an
orthogonal direct sum of three subspaces with special properties, viz.
\[
\cM^{(\vr)}=\cR^{(\vr)}\oplus\cH^{(\vr)}\oplus\cJ^{(\vr)},\qquad
\text{for\ }\vr\in\gD^\imag_+,
\lb{miss-tridec} \]
where $\cR^{(\vr)}$ consists of states belonging to lower-height
\fgA-multiplets and $\cH^{(\vr)} := [\cH_+,\cH_+]\cap\cM^{(\vr)}$
is spanned by multiple commutators of appropriate lower-height vacuum
vectors. What can we say about the remaining piece,
$\cJ^{(\vr)}$? Its states are vacuum vectors for \fgA, which cannot be
reached by multiple commutators inside the space of missing lowest
weight vectors, $\cH_+$. So a basis for $\cJ^{(\vr)}$ is part of a
basis for $\cH_+$. At the level of the Borcherds algebra \fgL, this
just means that the root $\vr$ is an imaginary simple root of
multiplicity $\dim\cJ^{(\vr)}$. For this reason we introduce the
so-called {\bf simple multiplicity} $\gm(\vr)$ of a root $\vr$ in the
fundamental Weyl chamber as
\[
\gm(\vr):= \dim\cJ^{(\vr)}.
\lb{def-simpmult} \]
Obviously we have $\gm (\vr) \leq \mult (\vr)$.
Once we know the simple multiplicity of a fundamental root, it is
clear how to proceed. Recursively by height, we adjoin to \fgA for
each fundamental root $\vr$ a set of $\gm(\vr)$ generators $\{e_j,
f_j, h_j\}$. This also explains why it is sufficient to concentrate on
fundamental roots. Indeed, by the action of the Weyl group we conclude
that the simple multiplicity of any non-fundamental positive imaginary
root is zero, while the Chevalley involution tells us that $\gm(\vr)=
\gm(-\vr)$ --- this just reflects that the fact that we adjoin the
Chevalley generators $e_j$ and $f_j$ always in pairs.

Let us point out that for ordinary (i.e.\ not generalized in the sense
of Borcherds) Kac--Moody algebras, for which {\em all} elements of any
root space are obtained as multiple commutators of the
Chevalley--Serre generators (by the very definition of a Kac--Moody
algebra!), we have $\gm(\vr)=0$, and therefore the notion of simple
multiplicity is superfluous.

\section{The Gnome Lie Algebra} \lb{sec:Gno}
The gnome Lie algebra $\Lie1$, which we will investigate in this
section, is the simplest example of a Borcherds algebra that can be
explicitly described as the Lie algebra of physical states of a
compactified string.  It is based on the lattice $\latt1$ as momentum
lattice of a fully compactified bosonic string in two space-time
dimensions. Since there aro no transversal degrees of freedom in $d=2$
and only longitudinal string excitations occur, the Lie algebra of
physical states may be regarded as the precise opposite of the fake
monster Lie algebra in 26 dimensions which has only transversal and no
longitudinal physical states. It constitutes an example of a
generalized Kac--Moody algebra which is almost ``purely Borcherds'' in
that with one exception, all its simple roots are imaginary
(timelike). The gnome Lie algebra is also a cousin of the true monster
Lie algebra because they both have the same root lattice, $\latt1$. In
fact, we shall see that the gnome Lie algebra is a Borcherds
subalgebra not only of the fake monster Lie algebra but also of any
Lie algebra of physical states associated with a momentum lattice that
can be decomposed in such a way that it contains \latt1 as a
sublattice.

\subsection{The lattice $\latt1$}
We start by summarizing some properties of the unique two-dimensional
even unimodular Lorentzian lattice $\latt1$. It can be realized as
$$
\latt1
:= \Zn (\tfrac12;\tfrac12) \oplus \Zn (-1;1)
=\bigl\{(\cl/2-n;\cl/2+n)\|\cl,n\in\Zn\bigr\},
$$
where for the (Minkowskian) product of two vectors our convention is
$$
(x^1;x^0)\.(y^1;y^0) := x^1 y^1 - x^0 y^0.
$$
Alternatively, we will represent the elements of $\latt1$ in a light
cone basis, i.e., in terms of pairs $\lico{\cl,n}\in \Zn\oplus\Zn$
with inner product matrix $\bigl(\begin{smallmatrix} 0 & -1 \\ -1 & 0
\end{smallmatrix}\bigr)$, so that $\lico{\cl,n}^2= -2\cl n$.  The
lattice points are shown in Fig.\ \ref{fig-1} below.

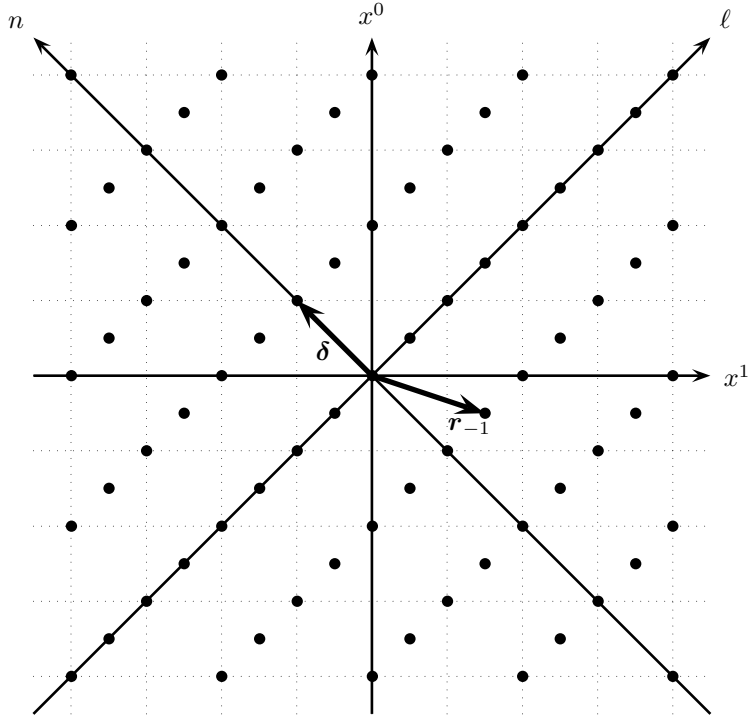
\begin{figure}[htb]
\begin{center}
\unitlength5mm
\begin{picture}(19,19)
\psset{xunit=5mm,yunit=5mm}
%
%
\multirput(1,0)(2,0){9}{\psline[linestyle=dotted,linewidth=.01pt](0,18)}
\multirput(0,1)(0,2){9}{\psline[linestyle=dotted,linewidth=.01pt](18,0)}
\psline[linewidth=1pt]{->}(0,9)(18,9)
\psline[linewidth=1pt]{->}(9,0)(9,18)
\psline[linewidth=1pt]{->}(0,0)(18,18)
\psline[linewidth=1pt]{->}(18,0)(0,18)
\uput[0](18,9){$x^1$}
\uput[90](9,18){$x^0$}
\uput[45](18,18){$\ell$}
\uput[135](0,18){$n$}
%
%
\psset{linecolor=black}
\multirput(1,1)(4,0){4}{
  \multirput(0,0)(0,4){4}{
    \multirput(0,0)(1,1){4}{\qdisk(-.225,0){.075}}}}
\multirput(1,17)(4,0){5}{\qdisk(0,0){.075}}
\multirput(17,1)(0,4){4}{\qdisk(0,0){.075}}
\psline[linewidth=2pt]{->}(9,9)(12,8)
\psline[linewidth=2pt]{->}(9,9)(7,11)
\psset{labelsep=2pt}
\uput[240](12,8){$\vrm$}
\uput[225](8,10){$\vgd$}
\end{picture}
\caption{The Lorentzian lattice \latt1 \lb{fig-1}}
\end{center}
\end{figure}

The main importance of this lattice for us derives from the fact
that it is the root lattice of the Lie algebra $\Lie1$ we are
about to construct. As already explained in the last section, allowed
physical string momenta have norm squared at most two and consequently
any root $\vgL$ for \Lie1 must obey $\vgL^2\leq 2$. There are no
lightlike roots here: the corresponding root spaces are
empty owing to the absence of transversal polarizations in two
dimensions. Therefore, imaginary roots for \Lie1 are all lattice
vectors lying in the interior of the lightcone. Real roots satisfy
$\vgL^2 =2$, and the lattice $\latt1$ possesses only two such
roots $\vgL =\pm \vrm$, where
$$
\vrm := (\tfrac32;-\tfrac12) = \lico{1,-1}.
$$
Our notation has been chosen so as to make explicit the analogy with
\0 where $\vrm$ is the over-extended root.
In addition we need the lightlike vector
$$
\vgd := (-1;1) = \lico{0,1},
$$
obeying $\vrm\cdot\vgd =-1$. Hence it serves as a lightlike Weyl
vector for $\Lie1$.\footnote{It is, however, only a ``real'' Weyl
vector since it has scalar product -1 with all real simple roots,
whereas it will not have the correct scalar products with all
imaginary simple roots. In fact, there is no true Weyl vector for
\Lie1.} It is analogous to the null root of the affine subalgebra $\9
\subset \0$, but the crucial difference is that for $\latt1$ it is
{\em not} a root (see above remark). Nonetheless, we can use $\vgd$ to
introduce the notion of level (again by analogy with $\0$), namely, by
assigning to a root $\vgL$ the integer
$$
\cl := -\vgd\X\vgL.
$$
This gives us a $\Zn$-grading of the set of roots. The
reflection symmetry of the lattice, which gives rise to the Chevalley
involution of \Lie1 and which introduces the splitting of the set of
roots into positive and negative roots, apparently changes the level
into its negative. Consequently, the sign of the level of a root
determines whether it is positive or negative, and for an analysis of
\Lie1 it is sufficient to consider positive roots only. We conclude that
the set of positive roots for \Lie1 consists of the level-1 root $\vrm$
and the infinitely many lattice vectors lying inside the forward
lightcone.

The Weyl group of \latt1 is very simple: since we can only reflect
with respect to the single root $\vrm$, it has only two elements and
is thus isomorphic to $\Zn_2$ just like the Weyl group of the monster
Lie algebra \ct{Borc90}. On any vector $\vx\in\Rn^{1,1}$ it acts as
$\fw_{-1}(\vx):= \vx -(\vx\X\vr_{-1})\vr_{-1}$; in light cone
coordinates we have the simple formula
$$
\fw_{-1}\bigl(\lico{\cl,n}\bigr) = \lico{n,\cl}.
$$
Hence the forward lightcone is the union of only two Weyl chambers;
the fundamental Weyl chamber leading to our choice of the real simple
root has been shaded in Fig.\ \ref{fig-2}. It is given by
$$
\cC = \bigl\{\vx\in\Rn^{1,1} \| \vx^2\le0, \vx\X\vrm\le0,
                              \vx\X\vgd\le0\bigr\}.
$$
The imaginary positive roots inside $\cC$ will be called fundamental
roots. Combining the action of the Weyl group with the reflection
symmetry of the lattice, the whole analysis of \Lie1 is thereby
reduced to understanding the root spaces associated with
fundamental roots.

Obviously, $\vrm$ and $\vgd$ span \latt1, and thus any positive
level-\cl root can be written as
$$
\vgL = \cl \vr_{-1} + n \vgd = \lico{\cl,n-\cl},
$$
where $n>\cl>0$ because of $\vgL^2 = 2\cl(\cl-n)$.

As explained in \ct{GebNic95a}, the DDF construction necessitates
the introduction of fractional momenta which do not belong to the
lattice. We define
$$
\va_\cl := \cl\vrm + \left(\cl-\frac1\cl\right)\vgd,\qquad
\vkl := - \frac1\cl\vgd,
$$
such that we can write down the so-called DDF decomposition
\[
\vgL = \va_\cl -\left(1-\frac12\vgL^2\right)\vkl
\lb{DDFdeco-sl2} \]
for any positive level-\cl root $\vgL$. The tachyonic momenta
$\va_\cl$ lie on a mass shell hyperbola $\va_\cl^2 =2$ which has
been depicted as a dashed line in Fig.\ \ref{fig-2} below. This figure
also displays the intermediate points (as small circles) ``between the
lattice'' required by the DDF construction, and allows us to visualize
how the lattice becomes more and more ``fractionalized'' with
increasing level. We call vectors $\va_\cl- m\vkl$, $0\le m\le
-\tfrac12\vgL^2$, which are not lattice points fractional roots. Note
that fractional roots can only occur for $\cl>1$. We stress that the
physical states associated with these intermediate points are {\em
not} elements of the Lie algebra $\Lie1$, as their operator product
expansions will contain fractional powers.

\begin{figure}[htb]
\begin{center}
\unitlength5mm
\begin{picture}(18,18)
\psset{xunit=5mm,yunit=5mm}
%
%
\pspolygon*[linecolor=lightgray](9,3)(4,18)(0,18)(0,12)
\multirput(1,0)(2,0){9}{\psline[linestyle=dotted,linewidth=.01pt](0,18)}
\multirput(0,1)(0,2){9}{\psline[linestyle=dotted,linewidth=.01pt](18,0)}
\psline[linewidth=1pt]{->}(0,3)(18,3)
\psline[linewidth=1pt]{->}(9,0)(9,18)
\psline[linewidth=1pt]{->}(6,0)(18,12)
\psline[linewidth=1pt]{->}(12,0)(0,12)
\uput[0](18,3){$x^1$}
\uput[90](9,18){$x^0$}
\uput[45](18,12){$\ell$}
\uput[135](0,12){$n$}
%
%
\pscurve[linewidth=.5pt]%
        (13.12, 0)(12.46, 1)(12, 2)(11.83, 3)%
        (12, 4)(12.46, 5)(13.12, 6)(13.90, 7)%
        (14.75, 8)(15.63, 9)(16.55, 10)(18, 11.54)
%
%
\multirput(12,4)(-1,1){12}%
   {\pscircle[fillstyle=solid,fillcolor=white,linewidth=.1pt]{.05}} 
\multirput(12.667,5.333)(-.667,.667){19}%
   {\pscircle[fillstyle=solid,fillcolor=white,linewidth=.1pt]{.05}} 
\multirput(13.5,6.5)(-.5,.5){23}%
   {\pscircle[fillstyle=solid,fillcolor=white,linewidth=.1pt]{.05}} 
\multirput(14.4,7.6)(-.4,.4){26}%
   {\pscircle[fillstyle=solid,fillcolor=white,linewidth=.1pt]{.05}} 
\multirput(15.333,8.667)(-.333,.333){28}%
   {\pscircle[fillstyle=solid,fillcolor=white,linewidth=.1pt]{.05}} 
\multirput(16.286,9.714)(-.286,.286){29}%
   {\pscircle[fillstyle=solid,fillcolor=white,linewidth=.1pt]{.05}} 
\multirput(17.25,10.75)(-.25,.25){29}%
   {\pscircle[fillstyle=solid,fillcolor=white,linewidth=.1pt]{.05}} 
%
%
\psset{linecolor=gray}
\rput(1,11){\qdisk(0,0){.075}}
\multirput(1,7)(1,1){3}{\qdisk(0,0){.075}}
\multirput(1,3)(1,1){5}{\qdisk(0,0){.075}}
\multirput(3,1)(1,1){5}{\qdisk(0,0){.075}}
\multirput(7,1)(1,1){11}{\qdisk(0,0){.075}}
\multirput(11,1)(1,1){7}{\qdisk(0,0){.075}}
\multirput(15,1)(1,1){3}{\qdisk(0,0){.075}}
%
%
\psset{linecolor=black}
\multirput(8,6)(1,1){10}{\qdisk(0,0){.075}}
\multirput(6,8)(1,1){10}{\qdisk(0,0){.075}}
\multirput(4,10)(1,1){8}{\qdisk(0,0){.075}}
\multirput(2,12)(1,1){6}{\qdisk(0,0){.075}}
\multirput(1,15)(1,1){3}{\qdisk(0,0){.075}}
\rput(12,2){\qdisk(0,0){.075}}
\end{picture}
\caption{Fundamental Weyl chamber, positive
         and fractional roots for \Lie1 \lb{fig-2}}
\end{center}
\end{figure}
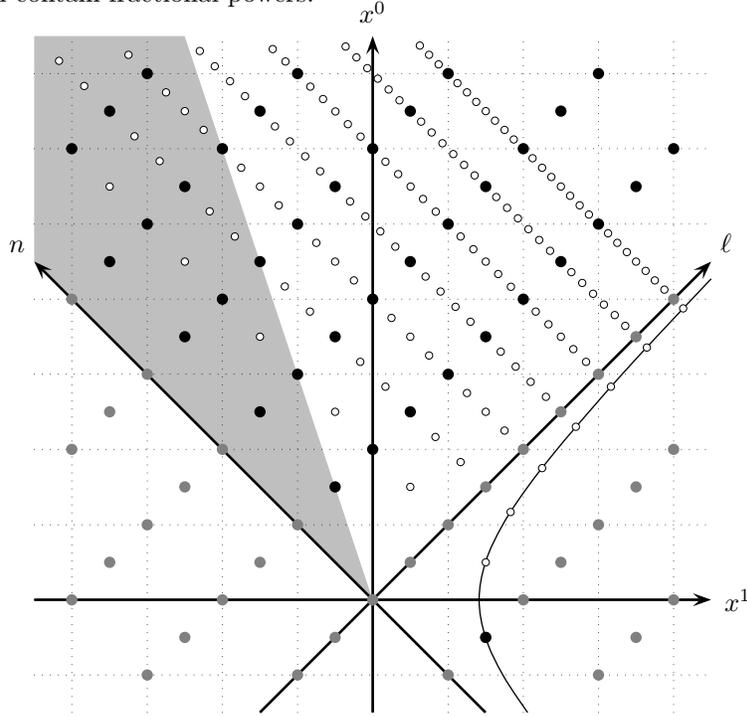

\subsection{Basic structure of gnome Lie algebra}
The gnome Lie algebra is by definition the Borcherds algebra $\Lie1$
of physical states of a bosonic string fully compactified on the
lattice $\latt1$.  We would first like to describe its root space
decomposition. To do so, we assign the grading $\lico{\cl,n}$ to any
string state with momentum $\lico{\cl,n}= \cl \vrm +(n-\cl)\vgd \in
\latt1$. The no-ghost theorem in the guise of Thm.\ \ref{thm-DDF} then
implies that the contravariant form $\cofo{\ |\ }$ is positive
definite on the piece of nonzero degree of the gnome Lie algebra
$\Lie1$. The degree $\lico{0,0}$ piece of $\Lie1$ is isomorphic to
$\Rn^2$, while the tachyonic states $\ket{\pm\vrm}$ yield two
one-dimensional subspaces of degrees $\lico{-1,1}$ and $\lico{1,-1}$,
respectively. With these conventions, the gnome Lie algebra looks
schematically like the monster Lie algebra (cf.\ \cite{Borc92})
\begin{figure}[htb]
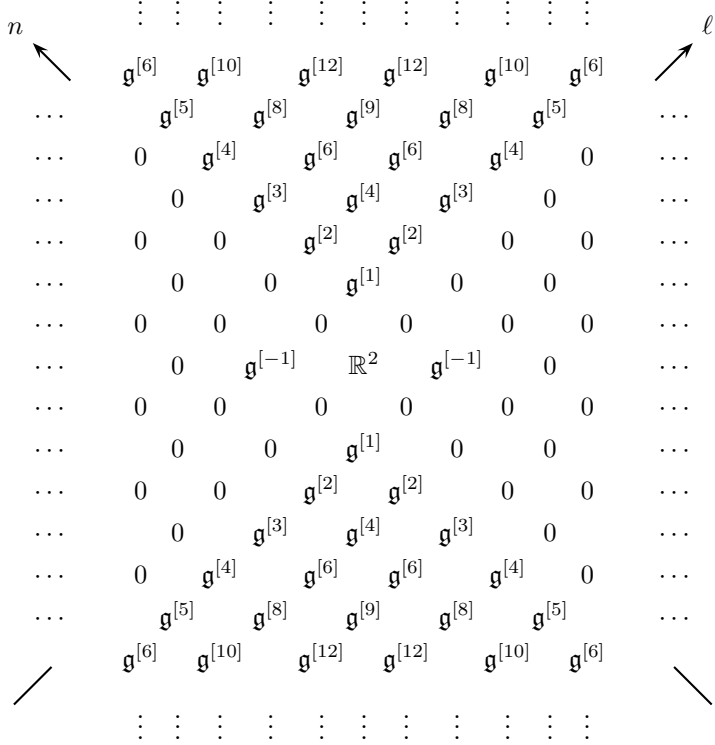

\begin{center}
\arraycolsep0pt
$$
\begin{array}{c@{\hspace{2em}}ccccccccccc@{\hspace{2em}}c}
       &\vdots&\vdots&\vdots&\vdots&\vdots&\vdots&%
        \vdots&\vdots&\vdots&\vdots&\vdots& \\[2ex]
\mbox{\psline{->}(.25,0)(-.25,.5)
\uput[135](-.25,.5){$n$}}
       &\fgn6& &\fgn{10}& &\fgn{12}&&\fgn{12}&  &\fgn{10}& &\fgn6&
       \mbox{\psline{->}(-.25,0)(.25,.5)
       \uput[45](.25,.5){$\cl$}} \\[.8ex]
\cdots &  &\fgn5&  & \fgn8  &   &\fgn9&   & \fgn8  &  &\fgn5&  &
        \cdots \\[.8ex]
\cdots & 0 &  &\fgn4&      &\fgn6&   &\fgn6&      &\fgn4&  & 0 &
        \cdots \\[.8ex]
\cdots &   & 0 &   & \fgn3  &   &\fgn4&   & \fgn3  &   & 0 &   &
        \cdots \\[.8ex]
\cdots & 0 &   & 0 &       &\fgn2&   &\fgn2&       & 0 &   & 0 &
        \cdots \\[.8ex]
\cdots &   & 0 &   &   0    &   &\fgn1&   &   0    &   & 0 &   &
        \cdots \\[.8ex]
\cdots & 0 &   & 0 &        & 0 &     & 0 &        & 0 &   & 0 &
        \cdots \\[.8ex]
\cdots &   & 0 &   &\fgn{-1}&   &\Rn^2&   &\fgn{-1}&   & 0 &   &
        \cdots \\[.8ex]
\cdots & 0 &   & 0 &        & 0 &     & 0 &        & 0 &   & 0 &
        \cdots \\[.8ex]
\cdots &   & 0 &   &   0    &   &\fgn1&   &   0    &   & 0 &   &
        \cdots \\[.8ex]
\cdots & 0 &   & 0 &       &\fgn2&   &\fgn2&       & 0 &   & 0 &
        \cdots \\[.8ex]
\cdots &   & 0 &   & \fgn3  &   &\fgn4&   & \fgn3  &   & 0 &   &
        \cdots \\[.8ex]
\cdots & 0 &  &\fgn4&      &\fgn6&   &\fgn6&      &\fgn4&  & 0 &
        \cdots \\[.8ex]
\cdots &  &\fgn5&  & \fgn8  &   &\fgn9&   & \fgn8  &  &\fgn5&  &
        \cdots \\[.8ex]
\psline{-}(0,0)(-.5,-.5)
       &\fgn6&&\fgn{10}& &\fgn{12}& &\fgn{12}& &\fgn{10}&&\fgn6&
       \psline{-}(0,0)(.5,-.5) \\[2ex]
       &\vdots&\vdots&\vdots&\vdots&\vdots&\vdots&%
        \vdots&\vdots&\vdots&\vdots&\vdots&
\end{array}
$$
\caption{Root space decomposition of the gnome Lie algebra \lb{fig-3}}
\end{center}
\end{figure}
\arraycolsep3pt
Here we have indexed the subspace associated with the root $\vgL=
\lico{\cl,n}$ by $[\cl n]$ because the dimension of this root space
depends only on the product $\cl n$. Indeed, since
$1-\tfrac12\lico{\cl,n}^2= 1+\cl n$ we have, according to
\Eq{genmult-1},
$$
\mult_{\Lie1}(\vgL)\equiv\dim\Lie1^{(\vgL)}=\pi_1(1+\cl n),
$$
where the partition function $\pi_1(n)$ was already defined in
\Eq{genmult-2}.


While this description of $\Lie1$ is rather abstract, we can
give a much more concrete realization of this Lie algebra by
means of the discrete DDF construction developed in \ct{GebNic95a}.
In fact, the DDF construction provides us with a \emph{complete basis}
for the gnome Lie algebra.

The single real simple root $\vrm$ of \latt1 gives rise to Lie algebra
elements (cf.\ Eq.\ \Eq{gen-def}
\[
h_{-1}:=\vrm(-1)\ket{\vO}, \qquad
e_{-1}:=\ket{\vrm}, \qquad
f_{-1}:=-\ket{-\vrm},
\lb{sl2} \]
which generate the finite Kac--Moody subalgebra $\fgA =
\fsl2 \equiv A_1\subset\Lie1$. On the other hand, there are
infinitely many imaginary (timelike) roots inside the lightcone. We
shall see that out of these all fundamental roots (except for one)
will be simple roots as well.

We notice that the one-dimensional Cartan subalgebra $\fhA$ spanned by
$h_{-1}$ does not coincide with the two-dimensional Cartan subalgebra
$\fh_{\latt1}$. Hence we need to introduce the Lie algebra
$$
\hgA := \fsl2+\fh_{\latt1}
      = \fsl2\oplus\Rn\vgL_0
$$
by adjoining to \fsl2 the element
$$
\vgL_0 := (\vrm+2\vgd)(-1)\ket{\vO},
$$
which commutes with \fsl2 and therefore behaves like a central
charge (but notice that the affine extension of \fsl2 is {\em not}
a subalgebra of \Lie1). It may be regarded as a remnant of the
Cartan subalgebra of the hyperbolic extension of a zero-dimensional
(virtual) Lie algebra.

We see that in this example the Lie algebra $\fgA$ is too small
to yield a lot of information (the ``smallness" of $\fgA$ is due
to the absence of transversal physical string states in two dimensions).
Nonetheless, there are infinitely many purely longitudinal physical
states present which are of the form
\[
\AL{-n_1}(\va_\cl)\dotsm\AL{-n_N}(\va_\cl)\ket{\va_\cl},
\lb{DDF-sl2} \]
where $n_1\geq n_2\geq\dots\geq n_N\geq 2$ and the longitudinal DDF
operators $\AL{-n_a}$ are associated with a tachyon momentum $\va_\cl$
and a lightlike vector \vkl satisfying $\va_\cl\X\vkl= 1$. Of course,
not all of these string states belong to $\Lie1$; in addition, we must
require that (cf.\ Eq.\ \Eq{DDFdeco-sl2})
$$
\vgL := \va_\cl - M \vk_\cl
$$
is a root, i.e. $\vgL\in\latt1$ with $\vgL^2\leq 2$, so that
$$
M:= \sum_{j=1}^N n_j = 1 - \tfrac12 \vgL^2 \ge 0.
$$
In other words, given a root $\vgL = \cl \vrm + n \vgd$, a basis of
the associated root space $\Lie1^{(\vgL)}$ is provided by longitudinal
DDF states of the above form with total excitation number $M=\cl(n-\cl
)+1$. For momenta of the form $\va_\cl - m \vk_\cl$, $0\le m<M$, such
that $m-1$ is not a multiple of $\cl$, i.e., for fractional roots
``between the lattice points'' (cf.\ Fig.\ \ref{fig-2}), we obtain
``intermediate (physical) states'' which are not elements of the Lie
algebra $\Lie1$. In fact, they are not full-fledged states of the
string model under consideration but rather states of the
uncompactified string model.

It is clear that, apart from the subalgebra $\hgA$, all elements of
the gnome Lie algebra are associated with imaginary roots. And since
none of the longitudinal states can be obtained by multiple
commutation of elements of \fsl2, all of them are missing states. Thus
\[
\cM_+^{(\vgL)} = \Lie1^{(\vgL)}
 = \Span\bigl\{\AL{-n_1}(\va_\cl)\dotsm\AL{-n_N}(\va_\cl)\ket{\va_\cl}\,
               \big|\, n_j>1, n_1+\dots+n_N= 1 - \tfrac12 \vgL^2\bigr\},
\]
for all $\vgL\in\gD^\imag_+$ and similarly for $\cM_-$. From the point
of view of \fsl2, all these states must be added ``by hand'' to fill
up \fsl2 to \Lie1. Having a complete basis for the space of missing
states the task is now to determine the complete set of imaginary
simple roots. In principle, this can be achieved in two steps. First,
we have to identify all the missing lowest weight vectors in
$\cM_+$. Then we have to determine a basis for the Lie algebra of
lowest weight vectors. This provides us with the complete information
about the imaginary simple roots and their multiplicities. In the next
subsection, this strategy is discussed in more detail and is illustrated
by some examples.

For the gnome Lie algebra, the information about the imaginary simple
roots and their multiplicities can be determined by means of the
Weyl--Kac--Borcherds denominator formula. One reason for this is the
simplicity of the Weyl group of \fsl2 which simplifies the denominator
formula enormously. It reads
\[
\left(x^{-1}-y^{-1}\right) \prod_{\cl,n>0}
  (1-x^\cl y^n)^{\pi_1(1+\cl n)}
&=& \bigg( x^{-1} - \sum_{n\ge\cl>0} \gm_{\cl,n} x^{\cl-1}y^n \bigg)
    - \bigl( x \longleftrightarrow y \bigr) \non
&=& \left(x^{-1}-y^{-1}\right)
    +\sum_{n\ge\cl>0}\gm_{\cl,n}
     \left(x^ny^{\cl-1}-x^{\cl-1}y^n\right),
\lb{denom-gnome} \]
where we write $x\equiv\re^{\lico{1,0}}$ and $y\equiv\re^{\lico{0,1}}$
for the generators of the group algebra of \latt1 and we put
$\gm_{\cl,n}\equiv \gm\bigl(\lico{\cl,n}\bigr)$. Recall that the
action of the Weyl group simply interchanges $x$ and $y$. Also note
that the fundamental roots have nonzero inner product with each other
so that there is no extra contribution of pairwise orthogonal
imaginary simple roots on the right-hand side. Therefore we are in the
fortunate situation that the sum on the right-hand side runs only once
over the imaginary simple roots and that the relevant coefficients are
just the simple multiplicities. Furthermore, the associated Lie
algebra of lowest weight vectors, $\cH_+$, is a free Lie algebra,
which follows from Thm.\ \ref{thm-miss} due to the fact that there are
no lightlike roots (cf.\ \ct{Juri97}).

We summarize: a set of imaginary simple roots for the gnome Lie
algebra $\Lie1$ is given by the vectors $\{\lico{\cl,n}\|
n\ge\cl\ge1\}$, each with multiplicity $\gm_{\cl,n}$ which is the
coefficent of $x^ny^{\cl-1}$ in the left-hand side of Eq.\
\Eq{denom-gnome} as generating function.

Expanding the latter, one
readily obtains the results (see Fig.\ \ref{fig-4})
\[
\gm_{1,n}
  &=& \pi_1(1+n)
  \qquad \text{for\ }n\ge1 \non
\gm_{2,n}
  &=& \pi_1(1+2n)-\pi_1(2+n)
      -\tfrac12\pi_1(1+\tfrac{n}{2})\bigl[\pi_1(1+\tfrac{n}{2})-1\bigr]
      -\sum_{k=1}^{\left[\frac{n-1}2\right]}\pi_1(1+k)\pi_1(1+n-k)
  \qquad \text{for\ }n\ge2, \nn
\]
where we have defined $\pi_1(1+\tfrac{n}{2}):=0$ for any odd integer $n$.
%
\begin{figure}[htb]
\begin{center}
\begin{minipage}{8cm}
\begin{tabular}{c|rrrrrr}
\multicolumn{7}{c}{\fbox{\large$\gm_{\cl,n}$}} \\[2ex]
$\cl\big\backslash n$\!\! & 1 & 2 & 3 & 4 & 5 & 6 \\ \hline
  1  &   1 &   1 &   2 &   2 &   4 &   4 \\
  2  &     &   0 &   1 &   2 &   6 &  10 \\
  3  &     &     &   3 &   6 &  20 &  40 \\
  4  &     &     &     &   5 &  36 & 101 \\
  5  &     &     &     &     &  63 & 239 \\
  6  &     &     &     &     &     & 331
\end{tabular}
\end{minipage}
\begin{minipage}{8cm}
\begin{tabular}{c|rrrrrr}
\multicolumn{7}{c}{\fbox{\large$\mult\bigl(\lico{\cl,n}\bigr)$}} \\[2ex]
$\cl\big\backslash n$\!\! & 1 & 2 & 3 & 4 & 5 & 6 \\ \hline
  1  &   1 &   1 &   2 &   2 &   4 &   4 \\
  2  &   1 &   2 &   4 &   8 &  14 &  24 \\
  3  &   2 &   4 &  12 &  24 &  55 & 105 \\
  4  &   2 &   8 &  24 &  66 & 165 & 383 \\
  5  &   4 &  14 &  55 & 165 & 478 &1238 \\
  6  &   4 &  24 & 105 & 383 &1238 &3660
\end{tabular}
\end{minipage}
\caption{Multiplicity of imaginary simple roots vs.\ dimension of root
         spaces \lb{fig-4}}
\end{center}
\end{figure}
The first formula tells us that all level-1 longitudinal states are
missing states associated with imaginary simple roots; from the second
we learn that this is no longer true at higher level since
$\gm_{2,n}<\pi_1(1+2n)$ and consequently some of the associated states
can be generated by commutation of level-1 states. In fact, one
easily sees that not only does $\gm(\vgL)$ not vanish in general, and
hence all higher-level roots are simple with a certain multiplicity,
but also that $\gm(\vgL)< \mult(\vgL)$ at higher level. This
illustrates the point we have already made in the introduction and in
the past \ct{GebNic95a}: while generalized Kac--Moody algebras such as
the gnome may have a rather simple structure in terms of the DDF
construction, they are usually quite complicated to analyze from the
point of view of their root space decompositions. For hyperbolic
Kac--Moody algebras, the situation is precisely the reverse: the
simple roots can be read off from the Coxeter--Dynkin diagram, but the
detailed structure of the root spaces is exceedingly complicated.

Due to the complicated pattern of the imaginary simple roots and their
multiplicities, the approach of decomposing $\Lie1$ into multiplets of
\fsl2 seems to be not very fruitful. One reason for this is that \fsl2
is just ``too small'' to yield non-trivial information about the full
Lie algebra -- in stark contrast to the algebra $\Lie9$ whose
corresponding subalgebra $\fgA = \0$ is much bigger. Another reason,
which is not so obvious, comes from the observation that for
increasing level the dimensions of the root spaces grow much faster
than the simple multiplicities. This explains why additional imaginary
simple roots are needed at every level. There is a beautiful example
where this situation is rectified. The true monster Lie algebra
\cite{Borc92} is a Borcherds algebra which is based on the same
lattice \latt1 as root lattice; but the multiplicity of a root
$\lico{\cl,n}$ is given by $c(\cl n)$ (replacing $\pi_1(1+\cl n)$)
which is the coefficient of $q^{\cl n}$ in the elliptic modular
function $j(q)-744= \sum_{n\ge-1}c_nq^n= q^{-1}+196884q+\dots$.  In
\cite{Borc92}, Borcherds was able to determine a set of imaginary
simple roots and their simple multiplicities by establishing an
identity for the elliptic modular function which turned out to be
precisely the above denominator formula. In that example, the
imaginary simple roots are all level-1 vectors $\lico{1,n}$ ($n\ge1$),
each with multiplicity $c(n)$. Thus the simple multiplicities are
large enough so that the level-1 \fsl2 vacuum vectors can generate by
multiple commutators the full Lie algebra of missing lowest weight
vectors.

Even though the infinite Cartan matrix looks rather messy, the gnome
Lie algebra \Lie1 has now been cast into the form of a Borcherds
algebra in the sense of Def.\ \ref{def-B}. The next
step in the analysis would be the calculation of the structure
constants. Since we have exhibited an explicit basis of the algebra in
terms of the DDF states, this can be done in principle. Practically,
however, the calculations still have to performed by use of
the humble oscillator basis $\{\ga^\mu_m\}$, whereas we would prefer
to be able to calculate the commutators of DDF states in
a manifestly physical way, i.e., in a formalism based on the DDF
operators only. For the transversal DDF operators this problem was
solved recently \ct{GebNic97}. However, since we are dealing with
purely longitudinal excitations here, one would certainly have to consider
exponentials of longitudinal DDF operators. This is technically much
more delicate, since the operators do not form a Heisenberg algebra
but a Virasoro algebra. Let us also point out the evident relation
between the gnome Lie algebra and Liouville theory, which remains to
be understood in more detail.

\subsection{DDF states and examples} \lb{subsec:gnome-examples}
We will now perform some explicit checks and for some examples exhibit
the split of the root spaces into parts that can be generated by
commutation of low-level elements and the remaining states which must
be adjoined by hand, and whose number equals the simple multiplicity
of the root in question. Since the actual calculations are quite
cumbersome it is helpful to use a computer. We would like to emphasize
that these examples not only provide completely explicit realizations
of the Lie algebra elements, but also enable us to determine the
``structure constants,'' whereas for other Borcherds algebras (such as
the true or the fake monster Lie algebra), investigations so far have
been limited to the determination of root space multiplicities and the
modular properties of the associated partition functions.

It is natural to investigate the subspace $\cM_+$ of missing states of
the gnome Lie algebra recursively level by level:
\[
\cM_+ = \bigoplus_{\cl>0}\cM^{[\cl]},\qquad
\cM^{[\cl]}:= \bigoplus_{\substack{\vgL\in\gD^\imag_+ \\
                                   \vgL\.\vgd=-\cl}}\cM^{(\vgL)}_+.
\]
We observe that, already at level 1, we have an infinite tower of
missing states; indeed, the states
\[
\AL{-n_1}(\vrm)\dotsm\AL{-n_N}(\vrm)\ket{\vrm} \lb{miss1-sl2}
\]
span $\cM^{[1]}$. Adjoining these states to the algebra is therefore
tantamount to adjoining infinitely many imaginary simple level-1 roots
$\vrm + n\vgd=\lico{1,n-1}$ ($n>1$) with multiplicity
$\pi_1(n)$.\footnote{As already mentioned, there are no proper
physical states on the lightcone, i.e., with momenta proportional to
the lightlike vectors $\vgd = \lico{0,1}$ and $\vrm + \vgd =
\lico{1,0}$, since these would require transversal polarizations.}
Although this statement is evident, we would like to demonstrate
explicitly that these states are indeed lowest weight vectors for
irreducible \fsl2-modules. So let us consider the state $ v_{\vgL} :=
\AL{-n_1}(\vrm)\dotsm\AL{-n_N}(\vrm)\ket{\vrm}$, where $\vgL:=
\vrm+n\vgd$, $n:= \sum_{j=1}^N n_j>1$. Using the adjoint action in
\Lie1 and the formulas for \fsl2 given in \Eq{sl2}, we infer that
\[
h_{-1}(v_{\vgL})
  &=& (2-n)v_{\vgL}, \non
f_{-1}(v_{\vgL})
  &\propto& L_{-1}\ket{n\vgd} \equiv 0, \non
(e_{-1})^{1-\vrm\.\vgL}(v_{\vgL})
  &\propto& L_{-1}\ket{n(\vrm+\vgd)} \equiv 0. \nn
\]
Note that the last two relations (the lowest weight and the null
vector condition, respectively) follow from momentum conservation
(cf.\ Eq.\ \Eq{tach-vertop}) and the fact that physical string states
in two dimensions are bound to be null states. Hence $v_{\vgL}$ is
indeed a vacuum vector for an irreducible \fsl2-module with spin
$\tfrac12(n-2)$. These multiplets can be constructed by repeated
application of the raising operator $e_{-1}$ which each time increases
the level by one. Clearly, the higher-level states belong to
irreducible \fsl2-multiplets, but the structure quickly becomes rather
messy. As already mentioned, we have to decompose each missing root
space $\cM^{(\vgL)}_+$ into an orthogonal direct sum of three
subspaces with special properties: one consists of states belonging to
lower-level \fsl2-multiplets, the other is made up of appropriate
multiple commutators of lower-level vacuum vectors, and the rest comes
from states corresponding to imaginary simple roots. We will now
illustrate this pattern by a few examples.

So the question is which of the higher-level states can be generated
by multiple commutators of the missing level-1 states. As it turns
out we will have to add new states at each higher-level root, apart
from an exceptional level-2 root which we will exhibit below.

We have calculated the following commutators (by means of
\texttt{MAPLE V})
\begin{subequations}
\[
\com{\ket{\vrm},A_{-3}^-\ket{\vrm}}&=&
  A_{-3}^-\ket{\va_2}, \lb{comm-1} \\
\com{\ket{\vrm},A_{-4}^- \ket{\vrm}}&=&
  \left(-\frac38 A_{-3}^-A_{-2}^- -\frac58
  A_{-5}^-\right)\ket{\va_2}, \lb{comm-2} \\
\com{\ket{\vrm},A_{-2}^- A_{-2}^-  \ket{\vrm}}&=&
  \left(-A_{-3}^-A_{-2}^- +A_{-5}^- \right)\ket{\va_2}, \lb{comm-3} \\
\com{\ket{\vrm},A_{-5}^- \ket{\vrm}}&=&
  \left(\frac{35}{64} A_{-7}^- +\frac{7}{32} A_{-5}^- A_{-2}^-
  +\frac{5}{64} A_{-3}^- A_{-2}^- A_{-2}^- \right)
  \ket{\va_2}, \lb{comm-4} \\
\com{\ket{\vrm},A_{-3}^- A_{-2}^-\ket{\vrm}} &=&
  \left(-\frac{61}{128} A_{-7}^- + \frac14 A_{-4}^- A_{-3}^-
  + \frac{7}{64} A_{-5}^- A_{-2}^-
  +\frac{37}{128} A_{-3}^- A_{-2}^- A_{-2}^- \right) \ket{\va_2},
  \lb{comm-5} \\
\com{A_{-2}^- \ket{\vrm},A_{-3}^- \ket{\vrm}} &=&
  \left( -\frac{83}{128} A_{-7}^- + \frac14 A_{-4}^- A_{-3}^-
  + \frac{41}{64} A_{-5}^- A_{-2}^- -\frac{21}{128} A_{-3}^- A_{-2}^-
  A_{-2}^-\right)\ket{\va_2}. \lb{comm-6}
\]
\end{subequations}
where $\va_2 = 2\vrm + \tfrac32 \vgd$ is the tachyonic level-2 root.
Furthermore, we have adopted the convention from \ct{GebNic95a}
according to which the DDF operators are always understood to be the
ones appropriate for the states on which they act (i.e.\ $A_m^-
(\vrm)$ on the l.h.s. and $A_m^-(\va_2)$ on the r.h.s.).

The first commutator generates an element of the root space associated
with $\vgL= 2\vrm+ 3\vgd$. But since this space is one-dimensional,
$\mult(2\vrm+3\vgd)= \pi_1(3)= 1$, we infer that we do not need an
additional imaginary simple root here (recall that
$\mult(2\vrm+n\vgd)= \mult\langle 2,n-2\rangle = \pi_1(2n-3)$).
This is, of course, a rather trivial observation because $\lico{2,1}$
is not a fundamental root anyhow.

The next two commutators leading to states in the root space
associated with $\vgL= 2\vrm+ 4\vgd$ are already more involved.
By taking suitable linear combinations we obtain
$A_{-3}^-A_{-2}^-\ket{\va_2}$ and $A_{-5}^-\ket{\va_2}$, which, as one
can easily convince oneself, already span the full two-dimensional
root space, $\mult(2\vrm+4\vgd)= \pi_1(5)= 2$. Consequently, this root
space can be entirely generated by commutators of level-1 missing
states, which means that $\gm_{2,2}= 0$. This is the only root in the
fundamental Weyl chamber which is not simple.

Let us finally consider a generic example. The commutators
\Eq{comm-4}--\Eq{comm-6} give states with momentum $\vgL= 2\vrm+
5\vgd$. Note that the commutators \Eq{comm-4} and \Eq{comm-5} are
states of spin 3/2 \fsl2-modules built on the vacuum vectors
$A_{-5}^-\ket{\vrm}$ and $A_{-3}^-A_{-2}^-\ket{\vrm}$,
respectively. In the notations of the last section (see Eq.\
\Eq{miss-tridec}), they span the two-dimensional space $\cR^{(\vgL)}$,
whereas $\cH^{(\vgL)}$ is one-dimensional with basis element given by
the commutator \Eq{comm-6} of two level-1 vacuum vectors. By building
suitable linear combinations these states can be simplified somewhat;
in this way, we get the three linearly independent states
\[
\left(A_{-7}^- + \frac35 A_{-5}^- A_{-2}^-\right)\ket{\va_2},\qquad
\left(A_{-3}^-A_{-2}^-A_{-2}^- -\frac75 A_{-5}^-
A_{-2}^-\right)\ket{\va_2},\qquad
\left(A_{-4}^-A_{-3}^-   + \frac{16}{5} A_{-5}^-
A_{-2}^-\right)\ket{\va_2}.
\]
However, we know that the full root space has dimension $\pi_1(7)=4$,
generated by the longitudinal DDF operators $A_{-3}^- A_{-2}^- A_{-2}^-,
A_{-4}^- A_{-3}^-, A_{-5}^- A_{-2}^-, A_{-7}^-$. Hence
$\cJ^{(\vgL)}$ must be one-dimensional. Indeed, the physical state
$$
\left(- 2457413 A_{-7}^- + 1354090 A_{-5}^- A_{-2}^-
      - 1613422 A_{-4}^- A_{-3}^-
      + 157593  A_{-3}^- A_{-2}^- A_{-2}^- \right)\ket{\va_2}
$$
is orthogonal to the above three states and cannot be generated by
commutation. Hence it is a missing state which must be added by hand
to arrive at the total count of four. We conclude that
$2\vrm+5\vgd$ is an imaginary simple root with simple multiplicity
$\gm_{2,3}=1$.

Of course, these explicit results are in complete agreement with the
Weyl--Kac--Borcherds formula predicting $\gm_{2,2}=0$ and
$\gm_{2,3}=1$ (cf.\ Fig.\ \ref{fig-4}).

\subsection{Direct sums of lattices}
We conclude this section with a remark about direct sums of lattices
and how this translates into the associated Lie algebras of physical
states.

Suppose we have two lattices $\gL_1$ and $\gL_2$. Then the direct sum
$$
\gL := \gL_1 \oplus \gL_2
$$
enjoys the following properties (see e.g.\ \ct{MooPia95}):
\ben
\item[(i)] $\rank\gL=\rank\gL_1+\rank\gL_2$;
\item[(ii)] $\sgn\gL=\sgn\gL_1+\sgn\gL_2$;
\item[(iii)] $\det\gL=(\det\gL_1)(\det\gL_2)$;
\item[(iv)] $\gL$ is even iff both $\gL_1$ and $\gL_2$ are even ;
\een
where $\sgn$ denotes the signature of a lattice. For $\gL$ to be even
Lorentzian we shall therefore assume that $\gL_1$ is even Lorentzian
and $\gL_2$ is even Euclidean. For example, the root lattice of \0 can
be decomposed into a direct sum of the unique even selfdual Lorentzian
lattice \latt1 in two dimensions and the \8 root lattice. More
generally, we have
$$
\latt{8n+1}=\latt1\oplus\gG_{8n},
$$
where $\gG_{8n}$ denotes an even selfdual Euclidean lattice of rank
$8n$.\footnote{As is well known (see e.g.\ \ct{ConSlo93}), there
exists only one such lattice for $n=1$ (associated with \8), two for
$n=2$ (associated with $\8\oplus\8$ and $\mathrm{Spin}(32)/\Zn_2$,
resp.), and 24 for $n=3$ (the 24 Niemeier lattices with the famous
Leech lattice as one of them). For higher rank, an explicit
classification seems impossible. This is due to the explosive growth
of the number of even selfdual Euclidean lattices according to the
Minkowski--Siegel mass formula which, for example, gives us
$8\times10^7$ as a lower limit on the number of such lattices with
rank 32.}

We would like to answer the question how the Lie algebra of physical
states in $\cF_{\gL}:= \cF_{\gL_1}\otimes\cF_{\gL_2}$ is built up from
the states in $\cF_{\gL_1}$ and $\cF_{\gL_2}$. This amounts to
rewriting both $\cP^1_{\gL}$ and $L_{-1}\cP^0_{\gL}$ as direct sums of
tensor products of subspaces of $\cF_{\gL_i}$. Using the facts about
tensor products of vertex algebras \ct{FrHuLe93} and that
$\cF^h_{\gL_2}=0$ for $h<0$, we deduce that any state in
$\gp\in\cP^1_{\gL}$ is a finite linear combination of the form
$$
\gp=\sum_{h=0}^H\gp_1^{1-h}\otimes\gp_2^h,
$$
with $\gp_i^h\in\cF^h_{\gL_i}$ and satisfying
\[
\gp_1^{1-h}\otimes L_{2,n}\gp_2^h
  &=& 0 \qquad \text{for\ }0\le h< n, \non
L_{1,n}\gp_1^{1-h}\otimes\gp_2^h+\gp_1^{1-h-n}\otimes L_{2,n}\gp_2^{h+n}
  &=& 0 \qquad \text{for\ }0\le h\le H-n, \non
L_{1,n}\gp_1^{1-h}\otimes\gp_2^h
  &=& 0 \qquad \text{for\ }H-n< h\le H, \non
\]
for all $n>0$. We immediately see that $\gp_2^0\in \cP^0_{\gL_2}=
\Rn\ket{\vO}_2$ and $\gp_1^{1-H}\in\cP^{1-H}_{\gL_1}$, but it is
difficult to extract from the above relations similar information
about the other states. Nonetheless, the last two observations are
sufficient to pinpoint the gnome Lie algebra inside \fgL. Namely, by
considering the special case $\gp= \gp_1^1\otimes\ket{\vO}_2$, we can
immediately infer that $\Lie1\cong \Lie1\otimes \ket{\vO}_2\subset
\fgL$. So the gnome Lie algebra is a Borcherds subalgebra of any Lie
algebra of physical states for which the root lattice can be
decomposed into a direct sum in such a way that \latt1 arises as a
sublattice. This in particular holds for the Lie algebras based on the
lattices \latt9, \latt{17}, and \latt{25}, respectively, the latter
being the celebrated fake monster Lie algebra \ct{Borc90}.

We can explore the decomposition of $\cP^1_{\gL}$ further by the use
of the DDF construction. Let us suppose that $\gL_1$ is the lattice
\latt1 and that $\gL_2$ has rank $d-2$ ($>0$). We shall write vectors
in $\gL$ as $(\vr,\vv)$, where $\vr\in \latt1$ and $\vv\in \gL_2$,
respectively, so that $(\vr,\vv)^2= \vr^2+ \vv^2$. We wish to find a
tensor product decomposition of the subspace of $\cP^1_{\gL_1}$ which
has fixed momentum component $\vr\in \gL_1$, i.e., of the space
$$
\cP^{1,\vr}_{\gL}
 := \cP^1_{\gL}\cap\bigoplus_{\vv\in\gL_2}\cF^{(\vr,\vv)}_{\gL}.
$$
The idea is to perform the DDF construction in a clever way such that
the $d-2$ transversal directions all belong to the Euclidean lattice
$\gL_2$ and thus the transversal DDF operators can be identified with
the string oscillators in $\cF_{\gL_2}$. We start from the DDF
decomposition $\vr= \va_\cl -\left(1-\frac12\vr^2\right)\vkl$ (see
Eq.\ \Eq{DDFdeco-sl2}), which gives rise to the decomposition
$$
(\vr,\vv)=\bigl(\va_\cl-\tfrac12\vv^2\vkl,\vv\bigr)
          -\bigl(1-\tfrac12(\vr,\vv)^2)\bigr(\vkl,\vO)
$$
within $\gL$. A suitable set of polarization vectors is obtained from any
orthonormal basis $\{\vgx^i|1\le i\le d-2\}$ of $\Rn\otimes_\Zn\gL_2$
by putting $\vgx^i\equiv(\vO,\vgx^i)$. From Thm.\ \ref{thm-DDF} it
follows that
$$
\cP^{1,\vr}_{\gL}=
\Span\bigl\{\AI{i_1}{-m_1}\dotsm\AI{i_M}{-m_M}
            \AI{-}{-n_1}\dotsm\AI{-}{-n_N}
            \ket{\va_\cl-\tfrac12\vv^2\vkl,\vv}\,\big|\,
            \vv\in\gL_2, m_1+\dots+n_N=1-\tfrac12(\vr,\vv)^2\bigr\}.
$$
For fixed $h:= \frac12\vv^2+ \sum_a m_a$, we may identify
$$
\AI{i_1}{-m_1}\dotsm\AI{i_M}{-m_M}\ket{\va_\cl-h\vkl,\vv}
\cong \ket{\va_\cl-h\vkl}_1\otimes
      \ga^{i_1}_{-m_1}\dotsm\ga^{i_M}_{-m_M}\ket{\vv}_2,
$$
or
$$
\Span\bigl\{\AI{i_1}{-m_1}\dotsm\AI{i_M}{-m_M}\ket{\va_\cl-h\vkl,\vv}\bigr\}
\cong \ket{\va_\cl-h\vkl}_1\otimes\cF^h_{\gL_2}.
$$
If we finally use the fact that $\cP^h_{\gL_1}$ for any integer $h$ is
generated by longitudinal operators we conclude that
$$
\cP^{1,\vr}_{\gL}\cong
\bigoplus_{h=0}^{1-\frac12\vr^2}\cP^{1-h,(\vr)}_{\gL_1}
                         \otimes\cF^h_{\gL_2}
$$
for any $\vr\in\gL_1$. There is one subtlety here concerning the
central charge. The longitudinal Virasoro algebra occurring on the
right-hand side as spectrum-generating algebra for any $\cP^h_{\gL_1}$
does not have the naive central charge $c= 24$ (like for the gnome Lie
algebra) but rather $c= 26-d$, the extra contribution coming from the
transversal space $\gL_2$. So for $d= 26$ we get modulo null states
the trivial representation of the longitudinal Virasoro algebra and
hence $\fgL^{\vr}\cong \cF^{1-\frac12\vr^2}_{\gL_2}$ in agreement with
the literature \ct{Borc92}.

\section{Missing Modules for \0} \lb{sec:E10}
We now turn to the hyperbolic Kac--Moody algebra $\fgA = \0$,
which arises as the maximal Kac--Moody subalgebra of the Borcherds
algebra \Lie9 of physical states associated with a subcritical open
bosonic string moving in 10-dimensional space-time fully compactified
on a torus, so that the momenta lie on the lattice \latt9. As such,
it plays the same role for \Lie9 as \fsl2 did for the gnome Lie algebra,
but is incomparably more complicated. Again, the central idea to
split the larger algebra $\Lie9$ into \0 and its orthogonal complement
which can be decomposed into a direct sum of \0 lowest and highest
weight modules, respectively. Since the root lattice of \0 is
identical with the momentum lattice \latt9, there is no need to extend
\0 by outer derivations. Thus we start from
$$
\Lie9 = \0 \oplus \cM
$$
where the space of missing states $\cM$ decomposes as
$$
\cM = \cM_+\oplus\cM_-,\qquad
\cM_\pm=\bigoplus_{v\in\cH_\pm}\fU(\0)v;
$$
each of the (irreducible) \0 modules $\fU(\0)v$ is referred to as a
``missing module". To be sure, this decomposition still does not
provide us with an explicit realization of the \0 algebra since we
know as little about the \0 modules as about the \0 itself (see
\ct{FeFrRi93} for some recent progress). On the other hand, we do gain
insight by combining the unknown algebra and its unknown modules into
something which we understand very well, namely the Lie algebra of
physical states $\Lie9$ for which a basis is explicitly given in terms
of the DDF construction. Moreover, we will formulate a conjecture
according to which all higher-level missing states can be obtained by
commuting the missing states at level 1 whose structure is completely
known. Our explicit tests of this conjecture for the root spaces of
$\vgL_7$ and $\vgL_1$ constitute highly non-trivial checks, but of
course major new insights are required to settle the question for
higher levels. We should mention that the results of the previous
section immediately show that the conjecture fails for the gnome Lie
algebra $\Lie1$. As we have already pointed out, the \fsl2 module
structure of the missing states for $\Lie1$ is not especially
enlightening due to the ``smallness" of \fsl2. Here the situation is
completely different, because \0 and its representations are ``huge"
(even in comparison with irreducible representations of the affine \9
subalgebra!). If our conjecture were true it would not only take us a
long way towards a complete understanding of \0 but also provide
another hint that \0 is very special indeed. Conversely, it would also
allow us to understand the Borcherds algebra $\Lie9$ by exhibiting its
complete set of imaginary simple roots. In addition to the fake
monster, the true monster, and the gnome Lie algebra, this would be
the fourth example of an explicit realization of a Borcherds algebra.

\subsection{Basics of $\0$ }
As momentum lattice for the completely compactified string we shall
take the unique 10-dimensional even unimodular Lorentzian lattice
\latt9.  It can be defined as the lattice of all points
$\vx=(x_1,\ldots,x_{9}|x_0)$ for which the $x_\mu$'s are all in $\Zn$
or all in $\Zn+\frac12$ and which have integer inner product with the
vector $\vl=(\frac12,\ldots,\frac12\|\frac12)$, all norms and inner
products being evaluated in the Minkowskian metric
$\vx^2=x_1^2+\ldots+x_9^2-x_0^2$ (cf.\ \ct{Serr73}).

To identify the maximal Kac--Moody subalgebra of the Borcherds algebra
\Lie9 of physical string states we have to determine a set of real
simple roots for the lattice.  According to \ct{Conw83}, such a set is
given by the ten vectors $\vr_{-1}, \vr_0, \vr_1, \ldots, \vr_8$ in
\latt9 for which $r_i^2=2$ and $\vr_i\X\vgr=-1$ where the Weyl vector
is $\vgr=(0,1,2,\ldots,8|38)$ with $\vr^2=-1240$.\footnote{Note that
$\vgr$ fulfills all the requirements of a grading vector for \Lie9}
Explicitly,
$$ \begin{array}{l@{\quad=\quad(}r@,r@,r@,r@,r@,r@,r@,r@,r@{\|}l}
   \vr_{-1} & 0 & 0 & 0 & 0 & 0 & 0 & 0 & 1 & -1 & 0),  \non
   \vr_{ 0} & 0 & 0 & 0 & 0 & 0 & 0 & 1 & -1 & 0 & 0),  \non
   \vr_{ 1} & 0 & 0 & 0 & 0 & 0 & 1 & -1 & 0 & 0 & 0),  \non
   \vr_{ 2} & 0 & 0 & 0 & 0 & 1 & -1 & 0 & 0 & 0 & 0),  \non
   \vr_{ 3} & 0 & 0 & 0 & 1 & -1 & 0 & 0 & 0 & 0 & 0),  \non
   \vr_{ 4} & 0 & 0 & 1 & -1 & 0 & 0 & 0 & 0 & 0 & 0),  \non
   \vr_{ 5} & 0 & 1 & -1 & 0 & 0 & 0 & 0 & 0 & 0 & 0),  \non
   \vr_{ 6} & -1 & -1 & 0 & 0 & 0 & 0 & 0 & 0 & 0 & 0),  \non
   \vr_{ 7} & \2 & \2 & \2 & \2 & \2 & \2 & \2 & \2 & \2 & \2),  \non
   \vr_{ 8} & 1 & -1 & 0 & 0 & 0 & 0 & 0 & 0 & 0 & 0). \nn
\end{array} $$
These simple roots indeed generate the reflection group of \latt9.
The corresponding Coxeter--Dynkin diagram associated with the Cartan
matrix $a_{ij}:= \vr_i\X\vr_j$ looks as follows:
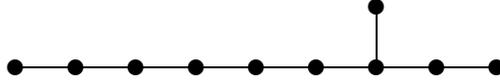
\[ \unitlength1mm
   \begin{picture}(66,10)
   \multiput(1,0)(8,0){9}{\circle*{2}}
   \put(49,8){\circle*{2}}
   \multiput(2,0)(8,0){8}{\line(1,0){6}}
   \put(49,1){\line(0,1){6}}
   \end{picture}  \nn  \]
The algebra \fgA is the hyperbolic Kac--Moody algebra \0,
defined in terms of generators and relations \Eq{gen-com}.
Moreover, from $|\det A|=1$ we infer that the root lattice $Q(\0)$
indeed coincides with \latt9, and hence $\hgA\equiv\fgA$ here.

The $\9$ null root is
$$
\vgd=\sum_{i=0}^8n_i\vr_i
       = (0,\ 0,\ 0,\ 0,\ 0,\ 0,\ 0,\ 0,\ 1\|1),
$$
where the marks $n_i$ can be read off from
$$
\left[\begin{array}{*{9}{c}}
      &   &   &   &    &    &  3 &   &   \\
     0& 1 & 2 & 3 &  4 &  5 &  6 & 4 & 2
      \end{array} \right]
$$
The fundamental Weyl chamber $\cC$ of $\0$ is the convex cone
generated by the fundamental weights $\vgL_i$,\footnote{Notice that
our convention is opposite to the one adopted in \ct{KaMoWa88}.  The
fundamental weights here are {\it positive} and satisfy $\vgL_i \X
\vr_j = - \gd_{ij}$.}
$$
\vgL_i = - \sum_{j=-1}^8 (A^{-1})_{ij} \vr_j
 \text{for $i=-1,0,1,\ldots 8$},
$$
where $A^{-1}$ is the inverse Cartan matrix. Thus,
$$
\vgL \in \cC    \Longleftrightarrow
     \vgL =     \sum_{i=-1}^8 k_i \vgL_i,
$$
for $k_i \in \Zn_+$. A special feature of $\0$ is that we need not
distinguish between root and weight lattice, since these are the same
for self-dual root lattices.\footnote{In the remainder, we will
consequently denote arbitrary roots by $\vgL$ and reserve the letter
$\vr$ for real roots.} Note also that the null root plays a special
role: the first fundamental weight is just $\vgL_{-1} = \vgd$, and all
null-vectors in $\cC$ must be multiples of $\vgL_{-1}$ since
${\vgL}_i^2 <0$ for all other fundamental weights.

We can employ the affine null root to introduce a \Zn-grading of \0. If
we introduce the so-called level \cl of a root $\vgL\in\gD(\0)$ by
$$
\cl := - \vgL \X \vgd,
$$
then we may decompose the algebra into a direct sum of subspaces of
fixed level, viz.
$$
\0=\bigoplus_{\cl\in\Zn}\0^{[\cl]},
$$
where
$$
\0^{[0]}\cong\9,\qquad
\0^{[\cl]}
  :=\bigoplus_{\substack{\vgL\in\gD(\0) \\
-\vgL\.\vgd=\cl}}\0^{(\vgL)}
\quad\text{for $\cl\ne0$}.
$$
Besides the obvious fact that $\cl$ counts the number of
$e_{-1}$ (resp.\ $f_{-1}$) generators in multiple commutators, the
level derives its importance from the fact that it grades the algebra
$\0$ with respect to its affine subalgebra $\9$ \ct{FeiFre83}.  The
subspaces belonging to a fixed level can be decomposed into
irreducible representations of $\9$, the level being equal to the
eigenvalue of the central term of the $\9$ algebra on this
representation (hence the full $\0$ algebra contains $\9$
representations of {\it all} integer levels!). Let us emphasize that
for general hyperbolic algebras there would be a separate grading
associated with every regular affine subalgebra, and therefore the
graded structure would no longer be unique.

Using the Jacobi identity it is possible to represent any subspace of
fixed level in the form
$$
\0^{[\cl]}=
 \underbrace{\bigl[\0^{[1]},\bigl[\0^{[1]},\ldots%
             \bigl[\0^{[1]},\0^{[1]}}_{\text{\cl times}}%
 \bigr]\bigr]\ldots\bigr],
$$
for $\cl>0$, and in an analogous form for $\cl<0$. This simple fact
turns out to be extremely useful in connection with the DDF
construction, as soon as one wishes to effectively construct
higher-level elements of \0.

Little is known about the general structure of this algebra. Partial
progress has been made in determining the multiplicity of certain
roots. Although the general form of the multiplicity formulas for
arbitrary levels appears to be beyond reach for the moment, the
following results for levels $\cl \leq 2$ have been established. For
$\cl =0$ and $\cl =1$, we have $\mult_{\0}(\vgL) = p_8(1-\2{\vgL}^2)$
(see \ct{Kac90}), i.e., the multiplicities are just given by the
number of transversal states; as was demonstrated in \ct{GebNic95a}
the corresponding states are indeed transversal. For $\cl =2$, it was
shown in \ct{KaMoWa88} that $\mult_{\0}(\vgL) = \xi(3-\2{\vgL}^2)$,
where $\sum_{n} \xi(n)q^n= \bigl[1- \phi(q^2)\big/
\phi(q^4)\bigr]\big/ \phi(q)^8$, $\phi(q)$ denoting the Euler function
as before.  Beyond $\cl=2$, no general formula seems to be known
although for $\cl=3$ the multiplicity problem was recently solved
\ct{BauBer97}.  However, the resulting formulas are somewhat implicit
and certainly more cumbersome than the above results. Of course, if
one is only interested in a particular root, the relevant multiplicity
can always be determined by means of the Peterson recursion formula
(see e.g.\ \ct{KMPS90}).

\subsection{Lowest and highest weight modules of \0}
We know from Thm.\ \ref{thm-miss} that $\cM_+$ (resp.\ $\cM_-$)
decomposes into a direct sum of lowest (resp.\ highest) weight
modules for \0. As before, $\cH_\pm$ denotes the subspace spanned by the
corresponding lowest and highest weight states, respectively.
Clearly, $\cH_\pm$ inherits from \Lie9 the grading by the level,
$$
\cH_\pm=\bigoplus_{\cl\gtrless0}\cHl,\qquad
\cHl:=\cH\cap\Lie9^{[\cl]}.
$$
Since the Chevalley involution provides an isomorphism between $\cHl$
and $\cHl[-\cl]$ and since we are ultimately interested in identifying
the imaginary simple roots and their multiplicities, it is sufficient
to restrict the explicit analysis to $\cH_+$. We will first study the
structure of the space $\cHl[1]$ and will explicitly demonstrate how
it is made up of purely longitudinal DDF states. Intuitively, this is
what one should expect. Recall that the level-1 sector of \0 is
isomorphic to the basic representation of $\9$ (cf.\ \ct{FeiFre83});
in terms of the DDF construction, it is generated by the transversal
states built on $\ket{\vrm}$, i.e., it is spanned by all states of the
form $\AI{j_1}{-m_1} \cdots \AI{j_k}{-m_k}\ket{\vrm}$ and their orbits
under the action of the \9 affine Weyl group \ct{GebNic95a}. Thus the
longitudinal states at level 1 do not belong to \0 and must be counted
as missing states. Furthermore, the level-1 transversal DDF operators
can be identified with the adjoint action of appropriate \9 elements
(corresponding to multiples of the affine null root). Hence the purely
longitudinal DDF states built on the level-1 roots of \0 are
candidates for missing lowest weight vectors. But apparently this set
can be further reduced, because each (real) level-1 root of \0 is
conjugated to some root of the form $\vrm+M\vgd$ ($M\ge0$) under the
action of the affine Weyl group. So we end up with purely longitudinal
states built on $\ket{\vrm}$ --- the same set we already encountered
in Sect.\ \ref{subsec:gnome-examples} for the case of the gnome Lie algebra!
And indeed, we have

\begin{Prp}  \lb{Prp-E10}
The space of missing level-1 lowest weight vectors consists of purely
longitudinal DDF states built on $\ket\vrm$,
$$
\cHl[1] =
 \Span\Big\{ \AL{-n_1}\dotsm\AL{-n_N} \ket{\vr_{-1}} \, \Big| \,
             n_1\geq n_2\geq\dots\geq n_N \geq 2 \Big\},
$$
i.e., it is (modulo null states) the longitudinal Virasoro--Verma
module with $\ket\vrm$ as highest weight vector.
In particular, $\vrm+n\vgd$ for any $n\ge2$ is an imaginary simple
root for \Lie9 with multiplicity $\gm(\vrm+n\vgd)=\pi_1(1+n)$.
\end{Prp}

\begin{proof}
Let us consider the state
$$
v_{\vgL} := \AL{-n_1}(\vrm)\dotsm\AL{-n_N}(\vrm)\ket{\vrm}
$$
with momentum $\vgL:= \vrm+M\vgd$, $M:= \sum_j n_j>1$. We first check
that, under the adjoint action in \Lie9, it is a lowest weight vector
for the basic representation of \9. Acting with either of the affine
Chevalley generators $e_i = \ket{\vr_i}$ and $f_i = -\ket{-\vr_i}$
($i=0,1,\dots,8$) on $v_{\vgL}$, we can move it through the
longitudinal DDF operators by the use of the general ``intertwining
relation'' \ct{GebNic97}
$$
E^{\vr}_n\AL{m}(\vrm)=\AL{m}(\va')E^{\vr}_n,
$$
where $\va':= \vrm+ \vr+ \vgd$ and $E^{\vr}_n$ denotes the step
operator associated with the real affine root $\vr+ n\vgd$. Thereby we
end up with the same state but where the Chevalley generator now acts
on $\ket\vrm$. The latter, however, is just a lowest weight vector
for the basic representation of \9, viz.
$$
f_i\ket\vrm=0, \qquad
e_i^{1-\vr_i\.\vrm}\ket\vrm=0, \qquad
\text{for } i=0,1,\dots,8\,,
$$
which is readily seen by inspection of the momenta. Indeed,
$(\vrm-\vr_i)^2= \bigl(\vrm+(1-\vr_i\X\vrm)\vr_i\bigr)^2=
2(2-\vr_i\X\vrm)\ge 4$ contradicting the mass shell condition
\Eq{mass-shell}. Since $h_i(v_{\vgL})= \vr_i\X\vgL v_{\vgL}=
-\gd_{i0}v_{\vgL}$, we conclude that $v_{\vgL}$ is a vacuum vector for
the adjoint action of \9 generating the basic
representation. Acccording to \ct{GebNic95a} it is given by the
transversal states built on $v_{\vgL}$, i.e., $\fU (\9)v_{\vgL}$ is
spanned by the states $\AI{j_1}{-m_1}\dotsm\AI{j_k}{-m_k}v_{\vgL}$,
where $\AI{j}{m}\equiv \AI{j}{m}(\vrm)$.

To show that the state $v_{\vgL}$ is a lowest weight vector for the
full \0 algebra, we have to check the remaining two Chevalley
generators. Again by momentum conservation, the state
$f_{-1}(v_{\vgL}) = -\bigl[\ket{-\vrm},v_{\vgL}\bigr]$ has momentum
$M\vgd$. But since the physical states associated with lightlike
momentum are purely transversal and are elements of \0, the resulting
missing state must be a null state (or vanishes identically). Within
\Lie9, we therefore have $f_{-1}(v_{\vgL})= 0$. On the other hand,
acting with the Chevalley generator $e_{-1}$ on $v_{\vgL}$ repeatedly,
say $k$ times, we obtain a state of momentum $\vgl= (1+k)\vrm
+M\vgd$. By the mass shell condition, this state identically vanishes
for $\vgl^2= 2(1+k)(1+k-M)> 2$, i.e., $k> M-1= 1- \vrm\X\vgL$. For $k=
M-1$, the momentum vector $\vgl$ is lightlike, and by the same
reasoning as before we conclude that the state is null also for this
value of $k$.

Altogether, we have shown that
$$
f_i(v_{\vgL}) = (e_i)^{1-\vr_i\.\vgL} (v_{\vgL})= 0 \qquad
\text{for } i=-1,0,1,\dots,8\,.
$$
These are the defining conditions for $v_{\vgL}$ to be a lowest weight
state for \0. Since $\ad h_i= \vr_i\X\vp$, it is clear that the lowest
weight is just $\vgL$. The fact that $f_{-1}$ annihilates the state
$v_{\vgL}$ in particular implies that we can ``only go up" in level (for
positive level lowest weight states) and that it is not possible to
cross the line $\cl=0$ by the action of $\0$.
\end{proof}

In the context of representation theory of hyperbolic Kac--Moody
algebras (see \ct{FeFrRi93}), the above result provides the first
examples of explicit realizations of unitary irreducible lowest weight
representations of the hyperbolic algebra \0. More specifically, they are
associated with lowest weights $\vgL_0+ m\vgL_{-1}$ for any
$m\ge0$. By commutation we even obtain an infinite set of missing
lowest-weight vectors with lowest weights $\cl\vgL_0+ n\vgL_{-1}$ for
any $\cl\ge1$ and $m\ge0$, on which we can built irreducible \0
modules. Analogous statements can be also made for other hyperbolic algebras
when we replace \latt9 by the root lattice of the hyperbolic
algebra. Due to the string realization this lattice should be even and
Lorentzian, conditions which rule out some hyperbolic algebras (see
e.g.\ \ct{sacl89} for a list of them).

The next question is now whether \Lie9 also provides realizations of
other lowest weight representations of \0. The results of the following
section suggest that this may not be the case. More specifically, we are
led to make the following

\begin{Cjt}
There are no imaginary simple roots for \Lie9 at level 2 or higher,
i.e., the Lie algebra of missing lowest weight states, $\cH_+$, is a free
algebra generated by the states given in Prop.\ \ref{Prp-E10}.
\end{Cjt}

Note that for the true monster Lie algebra the analogous claim is actually
valid: the imaginary simple roots are all of level 1. On the other hand,
the conjecture obviously fails for the gnome Lie algebra. The reason
for this is that the root spaces in the former example are much
bigger (due to the ``hidden'' extra 24 dimensions of the moonshine
module), even though the maximal Kac--Moody algebra in both examples
is the same, namely \fsl2. This appears to suggest that \0 has
just the right size so that the missing modules built on elements of
the free Lie algebra over \cHl[1] precisely fill up \0 to the full Lie
algebra of physical states.

At present, we are not aware of any convincing general argument in
favour of the above conjecture. In the next subsection, however, we
will verify it for two explicitly constructed non-trivial
root spaces. More specifically, we will consider a $201= 192+9$
dimensional and a $780= 727+53$ dimensional level-2 root space,
respectively, where the first contribution in each sum equals the dimension
of the \0 root space and the second term is the dimension of the space
of missing states. We will show for both examples that all the missing
states are contained in \0 modules built on level-1 missing lowest
weight vectors or on commutators of them. Of course, these two zeros
could be accidental like in the case of the gnome Lie algebra where
we also found a zero at level 2 (see Fig.\ \ref{fig-4}). In the latter
example, this was not unexpected since the root multiplicities in this
region of the fundamental cone are very low, anyway. For the \0 algebra,
by contrast, there is no apparent reason why all missing states
in certain level-2 root spaces should belong to \0 modules of the
conjectured type. The fact that they do in the cases we have studied
constitutes our primary motivation for the above conjecture.

\subsection{Examples: $\Lambda_7$ and $\Lambda_1$ }
We use the same system of polarizations vectors and DDF decomposition
as in \cite{GebNic95a}, which we recall here for convenience:
Explicitly, $\vgL_7$ is given by
\[ \vgL_7=\left[\begin{array}{*{9}{c}}
                &   &   &   &    &    &  7 &   &   \\
               2& 4 & 6 & 8 & 10 & 12 & 14 & 9 & 4
               \end{array} \right]
        =(0,\ 0,\ 0,\ 0,\ 0,\ 0,\ 0,\ 0,\ 0\|2),\nn \]
so $\vgL_7^2 = -4$. Its decomposition into two level-1
tachyonic roots is $\vgL_7=\vr+\vs+2\vgd$, where
\[ \vr:=\vr_{-1}
      &=&\left[\begin{array}{*{9}{c}}
                &   &   &   &    &    &  0 &   &   \\
               1& 0 & 0 & 0 &  0 &  0 &  0 & 0 & 0
               \end{array} \right]
        =(0,\ 0,\ 0,\ 0,\ 0,\ 0,\ 0,\ 1, -1\|0), \non
   \vs&:=&
         \left[\begin{array}{*{9}{c}}
                &   &   &   &    &    &  1 &   &   \\
               1& 2 & 2 & 2 &  2 &  2 &  2 & 1 & 0
               \end{array} \right]
       = (0,\ 0,\ 0,\ 0,\ 0,\ 0,\ 0, -1, -1\|0). \nn \]

Since $n=1-\frc12\vgL_7^2=3$ we have the DDF decomposition
$\vgL_7=\va-3\vk$ where $\vk:=-\frc12\vgd$ and
\[ \va:=\vr+\vs-\vk
       =(0,\ 0,\ 0,\ 0,\ 0,\ 0,\ 0,\ 0, -\frc32\|\frc12). \nn \]

As for the three sets of polarization vectors associated with the
tachyon momenta $\ket{\vr}$, $\ket{\vs}$ and $\ket{\va}$,
respectively, a convenient choice is
\[ \vgx_\ga&\equiv&\vgx_\ga(\vr)=\vgx_\ga(\vs)=\vgx_\ga(\va)
                 \mbox{ for } \ga=1,\ldots,7\ , \non
   \vgx_1&:=&(1,\ 0,\ 0,\ 0,\ 0,\ 0,\ 0,\ 0,\ 0\|0), \non
         &\vdots& \non
   \vgx_7&:=&(0,\ 0,\ 0,\ 0,\ 0,\ 0,\ 1,\ 0,\ 0\|0); \nn \\[.5em]
   \vgx_8(\vr)&:=&(0,\ 0,\ 0,\ 0,\ 0,\ 0,\ 0,\ 1,\ 1\|1), \non
   \vgx_8(\vs)&:=&(0,\ 0,\ 0,\ 0,\ 0,\ 0,\ 0, -1,\ 1\|1), \non
   \vgx_8\equiv\vgx_8(\va)
              &:=&(0,\ 0,\ 0,\ 0,\ 0,\ 0,\ 0,\ 1,\ 0\|0).\nn  \]
The little group is $\fW (\vgL_7 , \vgd ) = \fW (D_8 ) = S_8 {\mathbb
o} (\Zn_2 )^7$ of order $2^{14}3^1 5^1 7^1$.
We only have to evaluate the following commutator, where
$\epsilon$ denotes a cocyle-factor.
$$
\com{\ket{\vs},A_{-2}^- \ket{\vr}} =
\epsilon \bigg(-\frac12 A_{-3}^- -\frac{5}{6}A_{-1}^8 A_{-1}^8
  A_{-1}^8 +\frac{1}{3}A_{-3}^8 +\frac{1}{2}\sum_{\mu=1}^7 A_{-1}^\mu
  A_{-1}^\mu A_{-1}^8 \bigg)\ket{\va}
$$
To identify the remaining missing states, we act on this state with
the little Weyl group (which leaves the longitudinal contribution
invariant): $S_8$ permutes all transversal polarizations, and hence
generates another seven states. To see that the longitudinal state
can be separated from the transversal ones, we act with
$\fw_0 \cdots \fw_5 \fw_8 \fw_6 \fw_5 \cdots \fw_0$ on the above state;
this operation switches the relative sign between the transversal and the
longitudinal terms. Altogether we can thus isolate the following
nine states:
$$
\begin{array}{rc}
A_{-3}^-\ket{\va} \qquad &1 {\;\; \rm state}, \\[1ex]
\big\{2A_{-3}^i-8 A^i_{-1} A^i_{-1} A^i_{-1} + 3 A_{-1}^i \sum_{j=1}^8
A_{-1}^j A_{-1}^j\big\}\ket{\va} \qquad &8 {\;\; \rm states}.
\end{array}
$$
We use Roman letters $i,j$ running from 1 to 8 to label the
transversal indices. These nine states indeed span the orthogonal
complement of the 192-dimensional root space $\0^{(\vgL_7)}$ in
$\Lie{9}^{(\vgL_7)}$ as was already noticed in \cite{GeNiWe96} where
the result was derived by a completely different approach based on
multistring vertices and overlap identities.

Our second (more involved) example is the fundamental root $\vgL_1$ given by
\[ \vgL_1=\left[\begin{array}{*{9}{c}}
                &   &   &   &    &    &  9 &   &   \\
               2& 4 & 6 & 9 & 12 & 15 & 18 & 12 & 6
               \end{array} \right]
        =(0,\ 0,\ 0,\ 0,\ 0,\ 0,\ 1,\ 1,\ 1\|3), \nn \]
hence $\vgL_1^2=-6$ (our conventions used here are the same
as in \cite{BaeGeb97}). We have the DDF decomposition
$\vgL_1=\va - 4\vk$ where $\vk  = - \frc12 \vgd$ and
$$\va := \vgL_1 + 4 \vk =  \left(0,0,0,0,0,0,1,1,-1|1\right).$$
We will need two different decompositions of $\vgL_1$ into level-1
roots, namely:
\begin{enumerate}
\item $\vgL_1=\vr+\vs+3 \vgd$ with
\[ \vr:=\left[\begin{array}{*{9}{c}}
                &   &   &   &    &    &  0 &   &   \\
               1& 0& 0& 0 & 0 & 0 & 0 & 0 & 0
               \end{array} \right]
        =(0,\ 0,\ 0,\ 0,\ 0,\ 0,\ 0,\ 1,\ -1\|0),\nn \]
\[\vs:=\left[\begin{array}{*{9}{c}}
                &   &   &   &    &    &  0 &   &   \\
               1& 1& 0& 0 & 0 & 0 & 0 & 0 & 0
               \end{array} \right]
        =(0,\ 0,\ 0,\ 0,\ 0,\ 0,\ 1,\ 0,\ -1\|0),\nn \]
\item $\vgL_1=\vr'+\vs'+2 \vgd$ with
\[ \vr':=\left[\begin{array}{*{9}{c}}
                &   &   &   &    &    &  0 &   &   \\
               1& 1& 1& 0 & 0 & 0 & 0 & 0 & 0
               \end{array} \right]
        =(0,\ 0,\ 0,\ 0,\ 0,\ 1,\ 0,\ 0,\ -1\|0), \nn \]
\[\vs':=\left[\begin{array}{*{9}{c}}
                &   &   &   &    &    &  3 &   &   \\
               1& 1& 1& 3 & 4 & 5 & 6 & 4 & 2
               \end{array} \right]
        =(0,\ 0,\ 0,\ 0,\ 0,\ -1,\ 1,\ 1,\ 0\|1). \nn \]
\end{enumerate}
Although we will need several sets of polarization vectors adjusted
to these different decompositions, we will present the basis using
the following set, which is adjusted to the first decomposition:
\[
\vgx_\alpha &\equiv& \vgx_\alpha(\vr) = \vgx_\alpha(\vs)
=\vgx_\alpha(\va)\qquad \mbox{for } \alpha = 1, \ldots , 7,\nn \\
\vgx_1 & =&(1, \ 0, \ 0, \ 0, \ 0, \ 0, \ 0, \ 0, \ 0\|0),\nn\\
\vdots \non
\vgx_6 & =&(0, \ 0, \ 0, \ 0, \ 0, \ 1, \ 0, \ 0, \ 0\|0),\nn\\
\vgx_7 & =&\frc12 \sqrt{2}(0, \ 0, \ 0, \ 0, \ 0, \ 0, \ 1, \ 1, \
1\|1), \non
\vgx_8(\va) & =& \frc12 \sqrt{2}(0, \ 0, \ 0, \ 0, \ 0, \ 0, \ 1, \
-1, \ 0\|0),\non
\vgx_8(\vr) &=& \frc12 \sqrt{2}(0, \ 0, \ 0, \ 0, \ 0, \ 0, \ -1, \ 1,
\ 1\|1),\non
\vgx_8(\vs) &=& \frc12 \sqrt{2}(0, \ 0, \ 0, \ 0, \ 0, \ 0, \ 1, \ -1,
\ 1\|1).\nn
\]
The little Weyl group, $\fW(\vgL_1, \vgd)$, which is isomorphic to
$\Zn_2 \times \fW(E_7)$ in this case, acts on this set by permuting
$\vgx_1, \ldots \vgx_6$, as a $\Zn_2$ on $\vgx_8$ and by a more
complicated transformation on $\vgx_7$.  We worked out the following
commutator equations, $\epsilon$ denoting some irrelevant cocyle factor:
\begin{multline}
\com{\ket{\vs},A_{-3}^- \ket{\vr}} = \\
\begin{split}
\epsilon \bigg\{
&-\frac{1}{2}\sqrt{2}A_{-1}^8 A_{-3}^- +
  \frac{1}{8}\sum_{ \mu , \nu = 1}^{7} A_{-1}^\mu A_{-1}^\mu
  A_{-1}^\nu A_{-1}^\nu - \frac{3}{4}A_{-1}^8 A_{-1}^8 A_{-2}^- -
  \frac{3}{4}\sum_{\mu = 1}^{7} A_{-1}^\mu A_{-1}^\mu A_{-1}^8
  A_{-1}^8 \\
& - \frac{5}{24}A_{-1}^8 A_{-1}^8 A_{-1}^8 A_{-1}^8 -
  \frac{5}{6}A_{-1}^8 A_{-3}^8 + \frac{1}{4}\sum_{\mu = 1}^{7}
  A_{-1}^\mu A_{-1}^\mu A_{-2}^- + \frac{1}{2}\sum_{\mu = 1}^{7}
  A_{-1}^\mu A_{-3}^\mu \bigg\} \ket{\va},
\end{split}
\end{multline}
\begin{multline}
\com{\ket{\vs},A_{-1}^8 A_{-2}^- \ket{\vr}} = \\
\begin{split}
\epsilon \bigg\{
&\frac{7}{64}\sqrt{2}A_{-1}^8 A_{-1}^8 A_{-1}^8
  A_{-1}^8 - \frac{5}{32}\sqrt{2} \sum_{\mu = 1}^{7} A_{-1}^\mu
  A_{-1}^\mu A_{-1}^8 A_{-1}^8 + \frac{7}{16}\sqrt{2} A_{-1}^8
  A_{-3}^8 - \frac{1}{16}\sqrt{2}\sum_{\mu = 1}^{7} A_{-1}^\mu
  A_{-3}^\mu \\
& + \frac{1}{16}\sqrt{2}A_{-2}^- A_{-2}^- -
  \frac{1}{64}\sqrt{2}\sum_{ \mu , \nu = 1}^{7} A_{-1}^\mu A_{-1}^\mu
  A_{-1}^\nu A_{-1}^\nu - \frac{3}{4}A_{-1}^8 A_{-1}^8 A_{-2}^8 +
  \frac{1}{4}\sum_{\mu = 1}^{7} A_{-1}^\mu A_{-1}^\mu A_{-2}^8 -
  \frac{1}{4}\sqrt{2}A_{-4}^- \bigg\} \ket{\va},
\end{split}
\end{multline}
\begin{multline}
\com{ A_{-1}^8 \ket{\vs}, A_{-2}^- \ket{\vr}} = \\
\begin{split}
\epsilon \bigg\{
&-\frac{3}{4}A_{-1}^8 A_{-1}^8 A_{-2}^8 + \frac{1}{4}\sum_{\mu =
  1}^{7} A_{-1}^\mu A_{-1}^\mu A_{-2}^8 - \frac{1}{16}\sqrt{2}A_{-2}^-
  A_{-2}^- -\frac{7}{16}\sqrt{2}A_{-1}^8 A_{-3}^8 -
  \frac{7}{64}\sqrt{2}A_{-1}^8 A_{-1}^8 A_{-1}^8 A_{-1}^8 \\
&+\frac{1}{64}\sqrt{2}\sum_{ \mu , \nu = 1}^{7} A_{-1}^\mu A_{-1}^\mu
  A_{-1}^\nu A_{-1}^\nu
  + \frac{1}{16}\sqrt{2}\sum_{\mu = 1}^{7}
  A_{-1}^\mu A_{-3}^\mu + \frac{5}{32}\sqrt{2}\sum_{\mu = 1}^{7}
  A_{-1}^\mu A_{-1}^\mu A_{-1}^8 A_{-1}^8 +
  \frac{1}{4}\sqrt{2}A_{-4}^- \bigg\} \ket{\va},
\end{split}
\end{multline}
\begin{multline}
\com{A_{-1}^\mu\ket{\vs}, A_{-2}^- \ket{\vr}} = \\
\begin{split}
\epsilon \bigg\{
&-\frac{1}{4}\sum_{\nu = 1}^{7} A_{-1}^\nu A_{-1}^\nu A_{-2}^\mu +
  \frac{1}{2}\sqrt{2} A_{-1}^8 A_{-3}^\mu -
  \frac{1}{6}\sqrt{2}A_{-1}^\mu A_{-3}^8 - \frac{1}{2}A_{-1}^\mu
  A_{-3}^- + \frac{3}{4}A_{-1}^8 A_{-1}^8 A_{-2}^\mu \\
& +\frac{1}{12}\sqrt{2}A_{-1}^\mu A_{-1}^8 A_{-1}^8 A_{-1}^8 -
  \frac{1}{4}\sqrt{2} \sum_{\nu = 1}^{7} A_{-1}^\nu A_{-1}^\nu
  A_{-1}^\mu A_{-1}^8 \bigg\} \ket{\va},
\end{split}
\end{multline}
\begin{multline}
\com{\ket{\vs},A_{-1}^\mu A_{-2}^- \ket{\vr}} = \\
\begin{split}
\epsilon \bigg\{
&-\frac{3}{4}A_{-1}^8 A_{-1}^8 A_{-2}^\mu + \frac{1}{4}\sum_{\nu =
  1}^{7} A_{- 1}^\nu A_{-1}^\nu A_{-2}^\mu +
  \frac{1}{2}\sqrt{2}A_{-1}^8 A_{-3}^\mu - \frac{1}{6}\sqrt{2}
  A_{-1}^\mu A_{-3}^8 - \frac{1}{2}A_{-1}^\mu A_{-3}^- \\
& +\frac{1}{12}\sqrt{2}A_{-1}^\mu A_{-1}^8 A_{-1}^8 A_{-1}^8 -
  \frac{1}{4}\sqrt{2} \sum_{\nu = 1}^{7} A_{-1}^\nu A_{-1}^\nu
  A_{-1}^\mu A_{-1}^8 \bigg\} \ket{\va}.
\end{split}
\end{multline}
We need one more commutator, associated with a second DDF
decomposition. Namely,
\begin{multline}
\com{\ket{\vs}, A_{-2}^- \ket{\vr}} = \\
\begin{split}
\epsilon \bigg\{
&\frac{1}{32}\sum_{\mu = 1}^{7} A_{-1}^\mu A_{-1}^\mu A_{-1}^8
  A_{-1}^8 + \frac{1}{64}A_{-1}^8 A_{-1}^8 A_{-1}^8 A_{-1}^8 +
  \frac{1}{16}\sum_{\mu = 1}^{7} A_{-1}^\mu A_{-3}^\mu +
  \frac{1}{64}\sum_{ \mu , \nu = 1}^{7} A_{-1}^\mu A_{-1}^\mu
  A_{-1}^\nu A_{-1}^\nu \\
& - \frac{1}{16}A_{-2}^- A_{-2}^- -
  \frac{1}{3}\sqrt{3}A_{-1}^7 A_{-3}^- + \frac{1}{3}A_{-1}^6 A_{-3}^-
  \sqrt{6} + \frac{1}{6}\sqrt{2}A_{-3}^6 A_{-1}^7 +
  \frac{1}{4}\sqrt{2}A_{-1}^6 A_{-1}^7 A_{-1}^8 A_{-1}^8 \\
& -\frac{1}{3}\sqrt{2}A_{-1}^6 A_{-1}^7 A_{-1}^7 A_{-1}^7 +
  \frac{1}{6}\sqrt{2} A_{-1}^6 A_{-3}^7 + \frac{1}{16}A_{-3}^8
  A_{-1}^8 - \frac{1}{3}A_{-3}^6 A_{-1}^6 - \frac{1}{4}A_{-1}^6
  A_{-1}^6 A_{-1}^8 A_{-1}^8 \\
& - \frac{1}{6}A_{-3}^7 A_{-1}^7 - \frac{1}{8}A_{-1}^7 A_{-1}^7
  A_{-1}^8 A_{-1}^8 + \frac{1}{3}A_{-1}^6 A_{-1}^6 A_{-1}^6 A_{-1}^6 +
  \frac{1}{12}A_{-1}^7 A_{-1}^7 A_{-1}^7 A_{-1}^7 \\
& - \frac{1}{4}\sum_{\mu = 1}^{7} A_{-1}^\mu A_{-1}^\mu A_{-1}^6
  A_{-1}^6 - \frac{1}{8}\sum_{\mu = 1}^{7} A_{-1}^\mu A_{-1}^\mu
  A_{-1}^7 A_{-1}^7 + A_{-1}^6 A_{-1}^6 A_{-1}^7 A_{-1}^7 +
  \frac{1}{4}A_{-4}^- \\
& - \frac{2}{3}\sqrt{2}A_{-1}^6 A_{-1}^6 A_{-1}^6 A_{-1}^7 +
  \frac{1}{4}\sqrt{2}\sum_{\mu = 1}^{7} A_{-1}^\mu A_{-1}^\mu A_{-1}^6
  A_{-1}^7 \bigg\} \ket{\va}.
\end{split}
\end{multline}
We displayed this result using the basis of polarization associated
with the first decomposition. Appropriate linear combinations
and the little Weyl group action lead to the following 53 states,
spanning the orthogonal complement of the 727-dimensional root space
$\0^{(\vgL_1)}$ in $\Lie{9}^{(\vgL_1)}$.
We use the following conventions to label the transversal indices:
Roman letters $i,j,\dotsc$ run from 1 to 8, Greek letters from the
middle of the alphabet $\mu,\nu,\dotsc$ run from 1 to 7 and Greek
letters from the beginning of the alphabet $\alpha,\beta,\dotsc$ run
from 1 to 6.
$$
\begin{array}{rc}
A_{-1}^i A_{-3}^- \ket{\va} \qquad &
\text{8 states,} \\[1ex]
\big\{A_{-1}^8 A_{-1}^8 A_{-2}^i - \frac13 \sum_{\mu=1}^7
A_{-1}^\mu A_{-1}^\mu A_{-2}^i \big\}\ket{\va} \qquad &
\text{8 states,} \\[1ex]
\big\{A_{-2}^-A_{-2}^- - 4 A_{-4}^-\big\}\ket{\va}
      \qquad &
\text{1 state,} \\[1ex]
\big\{A_{-1}^\mu A_{-1}^8 A_{-1}^8 A_{-1}^8 - 3
\sum_{\nu=1}^7 A_{-1}^\nu A_{-1}^\nu A_{-1}^\mu A_{-1}^8 -2 A_{-1}^\mu
A_{-3}^8 \hphantom{\big\}\ket{\va},} & \\
+6 A_{-1}^8 A_{-3}^\mu\big\}\ket{\va} \qquad &
\text{7 states,} \\[1ex]
\big\{A_{-1}^\alpha A_{-3}^7 + A_{-1}^7 A_{-3}^\alpha - 2
A_{-1}^\alpha A_{-1}^7 A_{-1}^7 A_{-1}^7 -4 A_{-1}^\alpha
A_{-1}^\alpha A_{-1}^\alpha A_{-1}^7\hphantom{\big\}\ket{\va},} & \\
 + \frac32 \sum_{i=1}^8 A_{-1}^i A_{-1}^i A_{-1}^\alpha
A_{-1}^7 \big\}\ket{\va} \qquad &
\text{6 states,} \\[1ex]
\big\{A_{-3}^\alpha A_{-1}^\alpha -\frac{3}{2}A_{-3}^8
A_{-1}^8 + \frac{1}{2}A_{-3}^7 A_{-1}^7 - A_{-1}^\alpha A_{-1}^\alpha
A_{-1}^\alpha A_{-1}^\alpha\hphantom{\big\}\ket{\va},} &\\
 - \frac{1}{4}A_{-1}^7 A_{-1}^7 A_{-1}^7 A_{-1}^7
+\frac{3}{4}\sum_{i = 1}^{8} A_{-1}^i A_{-1}^i A_{-1}^\alpha
A_{-1}^\alpha\hphantom{\big\}\ket{\va},} &\\
 + \frac{3}{8}\sum_{i= 1}^{8} A_{-1}^i A_{-1}^i A_{-1}^7
A_{-1}^7 - A_{-1}^\alpha A_{-1}^\alpha A_{-1}^7 A_{-1}^7
\hphantom{\big\}\ket{\va},}&\\  + \frac{3}{8}\sum_{\mu =
1}^{7} A_{-1}^\mu A_{-1}^\mu A_{-1}^8 A_{-1}^8 - \frac{3}{8}A_{-1}^8
A_{-1}^8 A_{-1}^8 A_{-1}^8 \big\} \ket{\va}\qquad  &
\text{6 states,} \\[1ex]
\big\{ A_{-1}^\alpha A_{-3}^\beta + A_{-1}^\beta
A_{-3}^\alpha +\frac12 A_{-1}^\alpha A_{-1}^\beta A_{-1}^\beta
A_{-1}^\beta +\frac12 A_{-1}^\alpha A_{-1}^\alpha A_{-1}^\alpha
A_{-1}^\beta \hphantom{\big\}\ket{\va},} &\\
 -\frac32 \sum_{\substack{\gamma=1\\ \gamma \neq
\alpha,\beta}}^6 A_{-1}^\alpha A_{-1}^\beta A_{-1}^\gamma
A_{-1}^\gamma + \frac32 A_{-1}^\alpha A_{-1}^\beta A_{-1}^8 A_{-1}^8
\hphantom{\big\}\ket{\va},} &\\
 + \frac32 A_{-1}^\alpha A_{-1}^\beta A_{-1}^7 A_{-1}^7 + 4
A_{-1}^\gamma A_{-1}^\delta A_{-1}^\epsilon A_{-1}^\eta
\big\}\ket{\va}\qquad  &
\text{15 states,} \\[1ex]
\big\{\frac43 A_{-1}^8 A_{-3}^8 - \frac18 \sum_{\mu=1}^7
A_{-1}^\mu A_{-1}^\mu A_{-2}^- + \frac38 A_{-1}^8 A_{-1}^8
A_{-2}^-\hphantom{\big\}\ket{\va}}&\\
 +\frac13 A_{-1}^8 A_{-1}^8 A_{-1}^8 A_{-1}^8 + \frac14
\sum_{\mu=1}^7 A_{-1}^\mu A_{-1}^\mu A_{-1}^8 A_{-1}^8
\big\}\ket{\va}\qquad &
\text{1 state,} \\[1ex]
\big\{7 A_{-1}^8 A_{-3}^8 + \frac74 A_{-1}^8 A_{-1}^8
A_{-1}^8 A_{-1}^8 -\frac52 \sum_{\mu=1}^7 A_{-1}^ \mu A_{-1}^\mu
A_{-1}^8 A_{-1}^8 \hphantom{\big\}\ket{\va},}&\\
 -\frac14 \sum_{\mu,\nu=1}^7 A_{-1}^\mu A_{-1}^\mu
A_{-1}^\nu A_{-1}^\nu - \sum_{\mu=1}^7 A_{-1}^\mu A_{-3}^\mu
\big\}\ket{\va}\qquad &
\text{1 state.}
\end{array}
$$

\medskip
\noindent{\bf Acknowledgments:} H.N. would like to thank R.~Borcherds
for discussions related to this work.


\end{document}